\documentclass{aa}  
\usepackage{natbib}
\bibpunct{(}{)}{;}{a}{}{,} 

\newcommand{\Mm}{{\mathrm{\, Mm}}}
\newcommand{\iMm}{{\mathrm{\, Mm^{-1}}}}
\newcommand{\iiMm}{{\mathrm{\, Mm^{-2}}}}

\usepackage{graphicx}
\usepackage{txfonts}
\begin{document} 

\title{The effects of magnetic-field geometry on longitudinal oscillations of solar prominences: Cross-sectional area variation for thin tubes}
\titlerunning{Longitudinal oscillations with cross-sectional area variation}

\author{M. Luna \inst{1,2} \and A. J. D\'iaz \inst{3} \and R. Oliver \inst{3,4} \and J. Terradas \inst{3,4} \and J. Karpen \inst{5}}

\institute{Instituto de Astrof{\'{\i}}sica de Canarias, E-38205 La Laguna, Tenerife, Spain \and Universidad de La Laguna, Dept. Astrof{\'{\i}}sica, E-38206 La Laguna, Tenerife, Spain \and Departament F{\'i}sica, Universitat de les Illes Balears, E-07122 Palma de Mallorca, Spain \and Institute of Applied Computing \& Community Code (IAC$^3$), UIB, Spain \and NASA Goddard Space Flight Center, Greenbelt, MD 20771, USA}

\abstract{Solar prominences are subject to both field-aligned (longitudinal) and transverse oscillatory motions, as evidenced by an increasing number of observations.  Large-amplitude longitudinal motions provide valuable information on the geometry of the filament-channel magnetic structure that supports the cool prominence plasma against gravity. Our pendulum model, in which the restoring force is the gravity projected along the dipped field lines of the magnetic structure, best explains these oscillations. However, several factors can influence the longitudinal oscillations, potentially invalidating the pendulum model.
}{
The aim of this work is to study the influence of large-scale variations in the magnetic field strength along the field lines, i.e., variations of the cross-sectional area along the flux tubes supporting prominence threads.
}{
We studied the normal modes of several flux tube configurations, using linear perturbation analysis, to assess the influence of  different geometrical parameters on the oscillation properties.
}{
We found that the influence of the symmetric and asymmetric expansion factors on longitudinal oscillations is small.}{We conclude that the longitudinal oscillations are not significantly influenced by variations of the cross-section of the flux tubes, validating the pendulum model in this context.}


\maketitle

\section{Introduction}\label{sec:intro}
Solar prominences are subject to various types of oscillatory motions \citep[see review by][]{arregui2012}. In general these oscillations are a combination of movements transverse and longitudinal to the local magnetic field. In early magnetohydrodynamic (MHD) descriptions of longitudinal oscillations, the restoring force was proposed to be gas pressure gradients, as in the so-called slow modes. However, our  numerical simulations have shown that large-amplitude longitudinal oscillations (LALOs) are strongly influenced by the solar gravity projected along the dipped field lines that support the prominence \citep{luna2012b}. In fact, gravity is the most important restoring force under prominence conditions, according to our pendulum model, and the oscillation period depends mainly on the curvature of the field lines.  Subsequent parametric numerical studies of longitudinal oscillations for different tube geometries found good agreement with our pendulum model \citep{zhang2012,Zhang2013a}. In \citet{luna2012c}, hereafter called Paper I, we studied the effects of the field-line curvature on the longitudinal linear modes in flux tubes with uniform cross-sections (i.e., uniform magnetic field) but with different geometries. We found that slow modes become magnetoacoustic-gravity modes when dipped field lines are considered instead of horizontal straight lines. In those modes the restoring force is a combination of gravity projected along the field lines and gas pressure gradients, however, under typical prominence conditions the gravity dominates and is the main restoring force.

Here we extend the analysis of Paper I by including the variations of the magnetic field strength along the prominence-supporting flux tubes and determining their influence on the magnetosonic-gravity modes. These variations are converted to cross-sectional area variations according to conservation of magnetic flux: $B(s) \, A(s) = \mathrm{constant}$, where $B(s)$ and $A(s)$ are the magnetic field and cross-sectional area along the tube, and $s$ is the curved coordinate along the tube.  In the governing equations \citep[see, e.g.,][]{goossens1985,karpen2005}, derivatives of the magnetic field along the tube appear explicitly. Therefore longitudinal oscillations can be influenced by variations of the flux tube cross-section. This study establishes the effects of the flux tube cross-section on LALOs and the circumstances under which these effects are significant. In addition, this work contributes to prominence seismology because the results can be used to determine the magnetic field variations along the dipped field lines in observed prominence threads undergoing LALOs.

There are two main drawbacks to our model for the LALOs. We consider a flux tube as an isolated entity, and assume that the longitudinal oscillations are completely decoupled from the transverse oscillations. In reality, however, the flux tube is embedded in a large filament-channel structure, so the oscillations might depend on the global structure instead of the properties of isolated field lines. In addition the longitudinal and transverse motions might be coupled. However, studies of the linear global oscillations of a two-dimensional (2D) prominence structure revealed that, for the relatively small plasma-$\beta$ (ratio of gas pressure to magnetic pressure) typical for prominences, slow modes are decoupled from fast modes and are associated mainly with longitudinal motions  \citep{joarder1992,oliver1993}. \citet{goossens1985} studied oscillations of a 2.5D magnetic structure in a general situation of an arbitrary plasma-$\beta$. The authors determined the oscillatory spectrum of longitudinal, fast, and Alfv\'en modes. Part of both the longitudinal and Alfv\'en spectrum are formed of continuous modes. These continuous spectra are given by an eigenvalue problem on each magnetic surface and depend on the equilibrium quantities in different magnetic surfaces. This means that two different magnetic surfaces oscillate with different frequencies. In contrast, the discrete spectrum consists of modes of the whole structure where all the points oscillate with the same frequency. In general, both continua of Alfv\'en and longitudinal modes are coupled. However, the authors found that the two continua are not coupled if the magnetic structure is purely poloidal, i.e., the magnetic field is contained in the $xz$-plane and $B_y=0$. This is the situation considered in this work and justifies our treatment of the longitudinal oscillations in flux tubes as isolated structures. In addition, we are not interested in global modes or in the motions along the ignorable direction related to the Alfv\'en oscillations, rather we focus on the longitudinal continuum modes.

More recently, we studied nonlinear longitudinal and transverse motions in a 2D prominence structure \citep{luna2016}. We found that the longitudinal and transverse oscillations are not coupled, even in the nonlinear regime. The longitudinal motions agree very well with our pendulum model \citep{luna2012b}. In particular the plasma contained in each dipped flux tube oscillates with its own longitudinal frequency that depends mainly on the curvature of the field line dip, forming a continuum of frequencies. We also found that the transverse motions of the different field lines are identical, indicating that a global fast normal mode is established. This result confirms the validity of studying LALOs in prominences with the isolated flux tube of Paper I and the present work.

In Section \ref{sec:model} the model is presented, governing equations are described, and normal modes of the system are derived analytically for different flux-tube geometries. In Section \ref{sec:sym_tubes} the solutions for symmetric flux-tubes are shown. In Section \ref{sec:anti_tubes} the normal modes of asymmetric tubes are presented, whereas in Section \ref{sec:general_case} the mixed tubes are studied. Finally, in Section \ref{sec:conclusions} we summarize the results of this investigation and our conclusions.

\section{Model flux tube with area variation}\label{sec:model}

In this work we assume a low-$\beta$ plasma confined in a flux tube with a stationary magnetic field. In this regime the plasma motion is well described with the set of one-dimensional (1D) Equations (1)-(4) of \citet{karpen2005}. These equations are a simplified version of the equations described by \citet{goossens1985} under the assumption that the sound speed is much smaller than the Alfv\'{e}n speed. We additionally consider that the system is adiabatic with no heating and radiation terms. In contrast to Paper I, we include the variation of the cross-section of the flux tube. As we have discussed in \S \ref{sec:intro} there are no longitudinal normal modes of the global prominence structure under low-$\beta$ conditions; instead, the plasma on each field line oscillates independently. In this sense, we  use the terminology normal mode referring to the mode established on each field line instead of a global mode of the whole structure.

With these assumptions we linearize the modified \citet{karpen2005} set of equations to obtain the equations for the perturbed quantities
\begin{eqnarray}\label{linear-eq1}
\frac{\partial \rho_{1}}{\partial t} +\frac{1}{A} \frac{\partial A}{\partial s} v \rho_{0}+ v \frac{\partial \rho_{0}}{\partial s} + \rho_{0} \frac{\partial v}{\partial s} &=& 0 ~,\\ \label{linear-eq2}
\rho_{0} \frac{\partial v}{\partial t} + \frac{\partial p_{1}}{\partial s} - \rho_{1} g_{\parallel} &=& 0 ~,\\ \label{linear-eq3}
\frac{\partial p_{1}}{\partial t}+\frac{1}{A} \frac{\partial A}{\partial s} \gamma p_{0} v + v \frac{\partial p_{0}}{\partial s} + \gamma p_{0} \frac{\partial v}{\partial s} &=& 0~,
\end{eqnarray}
where $s$ is the coordinate along the flux tube, $A=A(s)$ is the cross-sectional area, $\rho$ is the density, $p$ is the gas pressure, $v$ is the perturbed velocity (a zero background velocity is considered), and $g_{\parallel}$ is the acceleration of gravity projected along the flux tube. The ``$0$'' index indicates the equilibrium quantity that depends on the coordinate $s$, while the ``$1$'' index denotes the perturbed quantity. The area expansion makes no contribution to the linearized momentum conservation (Eq. \ref{linear-eq2}) because the Lorentz force perturbation has no projection along the field lines. Additionally the unperturbed plasma is in hydrostatic equilibrium,
\begin{equation}\label{mechequil-eq}
\frac{\partial p_{0}}{\partial s} = \rho_{0} g_{\parallel}~.
\end{equation}
Combining Equations (\ref{linear-eq1})-(\ref{mechequil-eq}) we obtain the equation for the velocity perturbation
\begin{equation}\label{eq:wave-eq}
\frac{\partial^{2} v}{\partial t^{2}} - c_{s}^{2} \frac{\partial^{2} v}{\partial s^{2}} +F(s) \frac{\partial v}{\partial s} + G(s) v =0~,
\end{equation}\\
where $c_{s}^2=c_{s}^2(s)=\gamma p_{0}(s)/\rho_{0}(s)=\gamma \frac{k_B}{m_p} T(s)$. Functions $F(s)$ and $G(s)$ are
\begin{equation}\label{eq:f}
F(s) =-c^2_{s} ~\frac{\partial \ln A}{\partial s} - \gamma g_\parallel~,
\end{equation}
and
\begin{equation}\label{eq:g}
G(s)=-c^2_{s} ~\frac{\partial^2 \ln A}{\partial s^2} - \gamma \frac{\partial \ln A}{\partial s} g_\parallel - \frac{\partial g_\parallel}{\partial s}~.
\end{equation}
Assuming harmonic time dependence, $e^{i\omega t}$, the governing equation that describes the oscillatory motion of plasma embedded in a curved flux tube with cross-sectional area variations becomes
\begin{equation}\label{eq:gov_eq_full}
\frac{\partial^{2} v}{\partial s^{2}} -\frac{F(s)}{c_{s}^{2}} \frac{\partial v}{\partial s} + \left[ \frac{\omega^2}{c_{s}^{2} }- \frac{G(s)}{c_{s}^{2} }\right]  v =0~.
\end{equation}
This generalization of Equation (9) of Paper I is consistent with \cite{goossens1985}, \citet{terradas2013}, and \citet{kaneko2015}. These authors considered a more general situation with an arbitrary plasma-$\beta$. In their equations a tube speed defined as $c_T= v_A \, c_s / \sqrt{c_s^2+v_A^2}$ appears where $v_A$ is the local Alfv\'en velocity. In our approximation, $v_A \gg c_s$ and then $c_T \sim c_s$. In addition, in Equations (13) and (14) of \citet{terradas2013} derivatives $d \ln B /ds$ appear where $B(s)$ is the magnetic field strength. Considering tubes of constant magnetic flux, i.e., $B(s) A(s)=\mathrm{constant}$, we find that $d \ln A /ds=-d \ln B /ds$. With these considerations we find a correspondence between our Equations (\ref{eq:wave-eq})-(\ref{eq:g}) and the equations of the cited authors.

\subsection{Flux tube cross-section model}\label{sec:cross-section-discussion}
In general, the cross-section of a flux tube has a complex dependence of the position $s$ along the tube. For convenience, the cross-sectional area function is expanded as
\begin{equation}\label{eq:area}
A(s) = A_0 \, e^{ \epsilon_1 s + \epsilon_2 s^2 +...}= A_0 \, e^{ \sum_{p=1}^\infty \epsilon_p s^p}~,
\end{equation}
where $A_0=A(s=0)$. The origin of the $s$-coordinate, $s=0$, is selected at the central position of the tube. The $\epsilon_p$ factor quantifies the expansion or contraction terms of the tube. For ease, we call them expansion factors hereafter, having positive or negative values. It is important to note that we consider $A_0$ small enough that the bundle of field lines contained in the tube have almost identical characteristics. In this sense, all the plasma inside the tube oscillates with similar longitudinal frequencies and  the oscillations are in phase for a sufficiently long time.

\begin{figure}
\includegraphics[width=8cm]{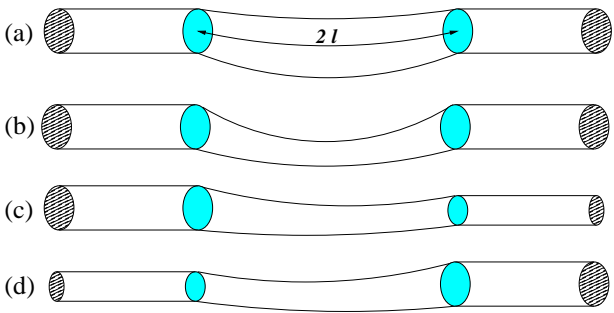}
\caption{Cartoon of the piecewise model considered with straight side coronal segments and curved central thread with four possible configurations: in (a) $\epsilon_1=0$ and $\epsilon_2 < 0$, in (b) $\epsilon_1=0$ and $\epsilon_2 > 0$, (c) in $\epsilon_1 < 0$ and $\epsilon_2=0$, and in (d) $\epsilon_1 > 0$ and $\epsilon_2=0$.}\label{fig:cartoon}
\end{figure}

We assume that the motion of the threads is around the central position of the dips. Then, we approximate the cross-sectional area by
\begin{equation}\label{eq:area-approx}
A(s) = A_0 \, e^{\epsilon_1 s + \epsilon_2 s^2}~.
\end{equation}
Considering the two terms in the exponent separately, the quadratic term yields a flux tube with a symmetric cross-section, $A(s)=A(-s)$ (see curved segment in Fig. \ref{fig:cartoon}(a) and \ref{fig:cartoon}(b)), while the linear term produces an asymmetric cross-section, i.e., neither symmetric nor antisymmetric (see curved tube segment in Fig. \ref{fig:cartoon}(c) and \ref{fig:cartoon}(d)). For this particular form of the exponential function, $A(s) \ne -A(-s)$ because $A(s)$ is always positive. For this reason, we call $\epsilon_2$ the symmetric expansion term and $\epsilon_1$ the asymmetric expansion term. 

A negative symmetric term produces flux tubes that contract with respect to the midpoint position (Fig. \ref{fig:cartoon}(a)). A positive symmetric term produces flux tubes that expand from the midpoint (Fig. \ref{fig:cartoon}(b)). In contrast, a negative asymmetric expansion term produces flux tubes that contract in the right-hand side of the tube and expand in the left-hand side of the tube (Fig. \ref{fig:cartoon}(c)); for a positive asymmetric expansion term the situation is reversed.  The positive and negative $\epsilon_1$ are equivalent. For example, a tube with negative $\epsilon_1$ is a tube with the same positive value of the expansion parameter that has flipped around $s=0$ (compare Figs. \ref{fig:cartoon}(c) and \ref{fig:cartoon}(d)). Therefore the normal modes for any tube with the same $|\epsilon_1|$ are identical.

The exponential dependence of the cross-sectional area of Equation (\ref{eq:area-approx}) gives two scale distances, $d_1=1/{|\epsilon_1 |}$ and $d_2=1/\sqrt{|\epsilon_2 |}$. These typical distances provide a more physical and intuitive way to measure the expansion or contraction of the flux tube: the position $s$ (measured from the midpoint) where the cross-sectional area has expanded or contracted $e$ times with respect to the central $A_0$. Thus, small $d_{1,2}$ involves strong expansion or contraction and large $d_{1,2}$ indicates weak expansion/contraction. For the extreme case of a tube with uniform cross-section, the typical distances are infinite.

\begin{figure}[!h]
\centering\includegraphics[width=9cm]{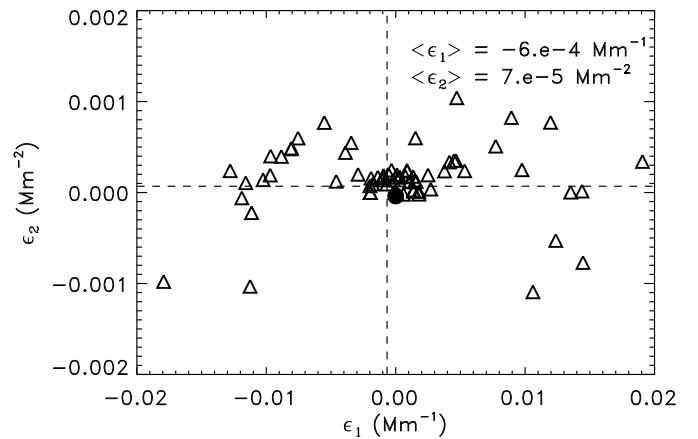}
\caption{Scatter plot of the flux tube expansion factors. Each point represents the asymmetric and symmetric factors $\epsilon_1$ and $\epsilon_2$, respectively, for the same flux tube. The flux tubes are from \cite{luna2012a} (triangles) and from \cite{terradas2013} (big dot). The vertical and horizontal dashed lines corresponds to the average values of $\epsilon_1$ and $\epsilon_2$ factors, respectively.\label{fig:epsilons}}
\end{figure}

Figure \ref{fig:epsilons} shows the fitted values of $\epsilon_1$ and $\epsilon_2$ for the flux tubes studied by \citet{luna2012a} and \citet{luna2012b}. In \cite{luna2012a} we investigated the process of formation of prominence condensation in $150$ flux tubes selected from a simulated 3D sheared-arcade magnetic structure \citep{devore2005}. This figure shows that $\epsilon_1$ is ten times larger than $\epsilon_2$. The asymmetric expansion factors $|\epsilon_1|$ are always $< 0.02 ~\mathrm{Mm^{-1}}$, indicating typical distances $d_1=1/\epsilon_1 > 50 \Mm$; the mean value has an associated expansion distance $d_1=300 \mathrm{~Mm}$. Similarly, the symmetric expansion factor is $|\epsilon_2| < 0.001\mathrm{~Mm^{-2}}$, indicating expansion distances $d_2=1/\sqrt{\epsilon_2} >32 \Mm$; the mean value is $d_2=120 \Mm$. The inequality between the mean typical expansion distances ($d_2 < d_1$) indicates that the symmetric expansion factor is more important than the asymmetric in the magnetic structure of \citet{luna2012a}. For the magnetic structure studied by \citet{terradas2013}, which is symmetric ($\epsilon_1 =0$), the typical expansion factor is $\epsilon_2 \sim -4 \times 10^{-5} \mathrm{~Mm^{-2}}$ (black dot in Figure \ref{fig:epsilons}). The corresponding expansion distance is $160 \mathrm{~Mm}$.

These expansion parameters delimit the possible range of values only for particular models, however, we estimated more general constraints on this range by considering the observed characteristics of cool prominence threads. High-resolution observations show that the threads have more or less constant widths along their lengths \citep[e.g.,][]{lin2003}, indicating that the expansion factors are small. Therefore the typical expansion distances should be larger than the thread length $2l$, such that $d_{1, \, 2} \gg 2 l$. In terms of the expansion factors this implies $|\epsilon_1| \ll 1/(2l)$ and $|\epsilon_2| \ll 1/(2l)^2$. Typical thread lengths are on the order of $10 \mathrm{~Mm}$, so $|\epsilon_1| \ll 0.1 ~\mathrm{Mm^{-1}}$ and $|\epsilon_2| \ll 0.01~\mathrm{Mm^{-2}}$. These limit values are extreme cases where the cross-sectional area differs by a factor of 3 between the thread ends. We expect lower values for real prominences.
 
\subsection{Isothermal plasma flux-tube segments} 
Solving Equation (\ref{eq:gov_eq_full}) is complicated because the sound speed, $c_s (s)$, is proportional to the square root of the plasma temperature, which depends on the position along the tube.  As in Paper I we adopt a simple approach and model the filament flux tube as a piecewise structure composed of three isothermal segments. The cool central thread has a constant sound speed of $c_s=c_{s p}$, while the two segments on either side are filled with hot  (1 MK) coronal plasma and have a constant sound speed of $c_s=c_{s c}$. Here $c_{s p}$ and $c_{s c}$ are typical prominence and coronal sound speeds, respectively. Additionally,  each isothermal segment has different radius of curvature and expansion factors. Figure \ref{fig:cartoon} illustrates the different tube models considered in this work.  As we found in Paper I, the shape of the coronal parts of the tube does not have a significant influence on longitudinal oscillations in prominences. Therefore we considered the simplest model for these side segments: straight, horizontal tubes with uniform cross-sections. The cool and dense prominence thread resides in the central segment. In our piecewise model, we assume that this central segment is dipped with uniform curvature and expansion factors but a nonuniform cross-section.

With the above considerations, Equations (\ref{eq:f}) and (\ref{eq:g}) can be written using Equation (\ref{eq:area-approx}) as
\begin{eqnarray}\label{eq:f_new}
F(s) &=&-c^2_{s} ~(\epsilon_1+2 \epsilon_2 s) - \gamma g_\parallel ~,\\ \label{eq:g_new}
G(s)&=&-2 c^2_{s} ~ \epsilon_2 - \gamma (\epsilon_1+2 \epsilon_2 s) g_\parallel - \frac{\partial g_\parallel}{\partial s}~.
\end{eqnarray}
In the side segments the two functions are identically zero, $F(s)=G(s)=0$. In contrast, we model the central segment as a circular arc with uniform radius of curvature $R$, where $s/R <<1$.
Thus, $g_\parallel \approx -g s/R$ and both functions can be approximated as
\begin{eqnarray}\label{eq:f_app}
F(s) &=&-c^2_{s} ~\epsilon_1+(-2 c^2_{s} \epsilon_2  + \gamma \frac{g }{R}) s ~,\\ \label{eq:g_app}
G(s)&=&-2 c^2_{s} ~ \epsilon_2 + \frac{g}{R} + \gamma \epsilon_1 \frac{g }{R} s+2 \gamma \epsilon_2  \frac{g }{R} s^2 ~.
\end{eqnarray}
In order to simplify the governing Equation (\ref{eq:gov_eq_full}), we define
\begin{eqnarray}\label{eq:norm_par1}
r&=& s \sqrt{\frac{\gamma g}{2 c_s^2 R}} ~, \\ \label{eq:norm_par2}
q_1&=&\frac{1}{\epsilon_1} \sqrt{\frac{\gamma g}{2 c_s^2 R}}~,\\ \label{eq:norm_par3}
q_2&=&\frac{1}{\epsilon_2} \frac{\gamma g}{2 c_s^2 R}~,
\end{eqnarray}
where the three quantities are dimensionless. With these definitions, Equation (\ref{eq:gov_eq_full}) can be written as,
\begin{eqnarray}\nonumber
\frac{d^2 v}{d r^2} + \left[\frac{1}{q_1} + 2 \left( \frac{1}{q_2} -1 \right) r \right] \frac{d v}{d r} + \\ \label{eq:adimensional_normalmode_equation}
\frac{2}{\gamma}\left[ \Omega^2 -1 +\frac{\gamma}{q_2}  -\frac{\gamma}{q_1}r -\frac{2 \gamma}{q_2} r^2 \right] v=0 ~,
\end{eqnarray}
where $\Omega = \omega/\omega_g$ and $\omega_g=\sqrt{g/R}$ is the pendulum frequency as defined in Paper I. The solutions of this Equation (\ref{eq:adimensional_normalmode_equation}) can be written in terms of confluent hypergeometric functions (CHFs) $M(a, b; x)$ \citep[see][]{abramowitz1972}, namely,
\begin{eqnarray}\nonumber 
v(r) = &C_1& e^{-r \left(\frac{r}{q_2}+\frac{1}{q_1}\right)}  \, M \left(-\lambda ;\frac{1}{2};\zeta ^2\right)+\\ \label{eq:normal_modes_solution}
&C_2& e^{-r \left(\frac{r}{q_2}+\frac{1}{q_1}\right)} \zeta~ \, M \left(-\lambda+\frac{1}{2} ;\frac{3}{2};\zeta ^2\right) ~,
\end{eqnarray}
where $C_1$ and $C_2$ are two arbitrary integration constants and 
\begin{eqnarray}
\lambda=\frac{q_2 \left(\Omega ^2-1\right)}{2 \gamma  \left(q_2+1\right)} ~,\\ \label{eq:zetadef}
\zeta(r)=\sqrt{\frac{q_2+1}{q_2}} r+\frac{1}{2 q_1 }\sqrt{\frac{q_2}{q_2+1}}~.
\end{eqnarray}
The general solution (Eq. (\ref{eq:normal_modes_solution})) is a linear combination of two kinds of solutions, which we call first-kind and second-kind solutions corresponding to the first and second terms of the equation. The first-kind solutions correspond to a situation where $C_2=0$, while the second-kind correspond to the $C_1=0$ case. The first kind has an even number of nodes (zeroes) of the velocity along the entire tube, whereas the second kind has an odd number of nodes.  In Paper I the solutions were classified as symmetric or antisymmetric according to their symmetry with respect to $s=0$ (the tube midpoint). The current first- and second-kind solutions are modulated by a factor $e^{-r \left(\frac{r}{q_2}+\frac{1}{q_1}\right)}$, however, so in general these solutions do not have well-defined symmetry with respect to the midpoint. Specifically, the solutions are likely to be asymmetric because the expansion factor $\epsilon_1\ne0$. When $\epsilon_1=0$, i.e. ,$q_1 \to \infty$, the solutions become symmetric or antisymmetric and the solutions for the uniform tube of Paper I are recovered. 

As we demonstrated in Paper I, Equation (\ref{eq:normal_modes_solution}) tends to a sinusoidal function in the case of a straight tube. The same result applies here in the hot tube segments with uniform cross-sectional area. The solutions of Equation (\ref{eq:adimensional_normalmode_equation}) in these two end segments are
\begin{equation} \label{eq:solution_hotsegments}
v(s)=D \, \sin  \left\{ \frac{\omega}{c_\mathrm{sc}} (L\pm s) \right\} \, 
,\end{equation}
where the $\pm$ symbol refers to the left and right tube segments solutions, respectively, $c_{s c}$ is the sound speed in the hot coronal plasma, and $D$ is an arbitrary constant.

\subsection{Boundary conditions}
The velocity solutions for each region of our piecewise flux-tube model must be joined by the boundary conditions. The resulting matched solutions are the normal modes of the system. To solve this problem we need two types of boundary conditions: no-flow conditions at the chromospheric footpoints of the tube (located at $s = \pm L$, where $2 L$ is the length of the supporting magnetic flux tube) and zero-gradient conditions at the plasma segment interfaces. First we consider the boundary condition at the interface of two different tube segments.  Equations (\ref{linear-eq1})-(\ref{linear-eq3}) yield
\begin{eqnarray}\label{eq:boundc1}
\left[ v \right] =0 ~,\\ \label{eq:boundc3}
\frac{1}{v}\left[ \frac{d v}{d s} \right] =- \frac{1}{A} \left[ \frac{d A}{d s} \right] = - \left[ \epsilon_1 +2  \epsilon_2  s_b \right]~,
\end{eqnarray}
where $\left[ a \right]=\lim_{\delta \to 0} a(s_b+\delta) - a(s_b-\delta) $ denotes the difference of magnitude $a$ on both sides of the boundary at $s=s_b$. In addition, the equilibrium cross-sectional area should be continuous in order to fulfill the governing equations: $\left[ A \right] =0$. These equations indicate that the velocity is continuous across each intersection of the segments, however the solutions are not necessarily derivable. Condition (\ref{eq:boundc3}) implies that the flow discharge, $A \, v$,  is derivable at the segment intersections. Because the two side segments consist of uniform tubes without expansion, the boundary condition can be written as
\begin{equation}\label{eq:bc_detailed_interphase}
\frac{1}{v(\pm l)}\left[ \frac{d v}{d s} \right]_{\pm l} =  \pm \left( \epsilon_1 +2  \epsilon_2  (\pm l) \right)~,
\end{equation}
where $s=\pm l$ are the positions of cool thread ends.

As in Paper I, we model the boundary condition at the ends of the flux tube as
\begin{equation}\label{eq:foot_bc}
v\left(s=\pm L\right)=0~,
\end{equation}
where we are assuming that the energy carried by the oscillating thread is not enough to perturb the lower, denser layers of the solar atmosphere. This assumption is probably valid for the linear regime addressed here, but may require modification for large-amplitude longitudinal oscillations.

\subsection{Normal mode solutions}

The solutions to this piecewise model are given by Equation (\ref{eq:normal_modes_solution}) in the central dense segment and Equation (\ref{eq:solution_hotsegments}) in both side segments, namely
\begin{equation}
v(s)=\left\{ \begin{array}{ll}
    B_{} \, \sin  \left\{ \frac{\omega}{c_\mathrm{sc}} (L+s) \right\}, &-L < s \le -l,\\
    \\
    C_{1} \, e^{-r \left(\frac{r}{q_2}+\frac{1}{q_1}\right)}  \, M \left(-\lambda ;\frac{1}{2};\zeta ^2\right)+\\ 
C_{2} \, e^{-r \left(\frac{r}{q_2}+\frac{1}{q_1}\right)} \zeta~ \, M \left(-\lambda+\frac{1}{2} ;\frac{3}{2};\zeta ^2\right), & -l < s \le l , \\
\\
    D_{} \, \sin  \left\{ \frac{\omega}{c_\mathrm{sc}} (L-s) \right\}, & l < s < L . \end{array} \right.
\label{eq:full_sol_guess}
\end{equation}
We have ensured that the no-flow boundary condition (\ref{eq:foot_bc}) is enforced by choosing a sine function in the hot regions that vanishes at $s=\pm L$.

Next, we apply the boundary conditions (\ref{eq:bc_detailed_interphase}) and (\ref{eq:foot_bc}) at the interface between the cool and hot plasma at $s=\pm l$ to Equation (\ref{eq:full_sol_guess}). In the following sections we present the resulting solutions derived for different tube geometries.

\section{Symmetric tubes with $\epsilon_1 = 0$ and $\epsilon_2\ne0$}\label{sec:sym_tubes}
\subsection{Eigenvalues}

In this section we assume that the tubes are symmetric with respect to the tube midpoint: $\epsilon_2 \neq 0$ and $\epsilon_1=0$ (Figs. \ref{fig:cartoon}(a) and \ref{fig:cartoon}(b)). In this case the first- and second-kind solutions are symmetric or antisymmetric with respect to $s=0$, so the dispersion relation can be factorized into two factors. Solving the eigenvalue problem for each factor, we obtain the symmetric and antisymmetric frequencies and eigenfunctions separately. 

In order to simplify the expressions we define a new set of dimensionless variables
\begin{equation}
\Omega=\frac{\omega}{\omega_{\mathrm{g}}},~W=\frac{l}{L}, ~\Phi=\frac{\omega_{\mathrm{g}} L}{c_{s c}}, ~\chi=\frac{c^{2}_\mathrm{sc}}{c^{2}_\mathrm{sp}},~\gamma_n=\gamma+\frac{2 \epsilon_2 c_\mathrm{sp}^2}{\omega_g^2}~ \,
\label{dim_units}
,\end{equation}
where $\omega_\mathrm{g}=\sqrt{\frac{g}{R}}$ is the pendulum frequency. With these definitions the dispersion relation for symmetric modes takes the form
\begin{eqnarray}\nonumber
&& \Omega \cot (\Phi  (1-W) \Omega ) =  \\ \label{eq:dr_sym_tube_1stkind}
 &&\Phi  \chi  W \left(\Omega ^2-1\right) \times \frac{ M \left(1+\frac{1- \Omega ^2}{2 \gamma_n  };\frac{3}{2};\frac{1}{2} W^2 \Phi ^2  \gamma_n  \chi \right)}{M \left(\frac{1 -
    \Omega ^2}{2 \gamma_n};\frac{1}{2};\frac{1}{2} W^2 \Phi ^2 \gamma_n  \chi \right)}~.
\end{eqnarray}
This dispersion relation is the generalization of the symmetric case without tube expansion (Eq. 22 of Paper I). This expression (\ref{eq:dr_sym_tube_1stkind}) transforms to the Paper I equation when $\gamma_n \to \gamma$. 

Similarly, the dispersion relation for the antisymmetric modes is
\begin{eqnarray} \nonumber
 \Omega  \cot (\Phi  (1-W) \Omega ) = && -\frac{1}{\Phi  W}-\frac{1}{3}\Phi  \chi  W \left(1+ \gamma_n-\Omega ^2\right) \times\\ \label{eq:dr_sym_tube_2ndkind}
&&  \frac{ M\left(\frac{3}{2}+\frac{1-\Omega^2}{2 \gamma_n};\frac{5}{2};\frac{1}{2} W^2 \Phi ^2 \gamma_n \chi \right)}{M\left(\frac{1}{2}+\frac{1-\Omega^2}{2 \gamma_n};\frac{3}{2};\frac{1}{2} W^2  \Phi ^2 \gamma_n \chi \right)} ~.
 \end{eqnarray}
It is important to note that there is a typographical error in the antisymmetric mode dispersion relation of Paper I (Eq. 23): a minus sign should appear on the right side of the equation. Taking this into account, Equation (\ref{eq:dr_sym_tube_2ndkind}) converges to the antisymmetric corrected dispersion relation of Paper I when $\gamma_n \to \gamma$.

\begin{figure}[!ht]
\includegraphics[width=0.5\textwidth]{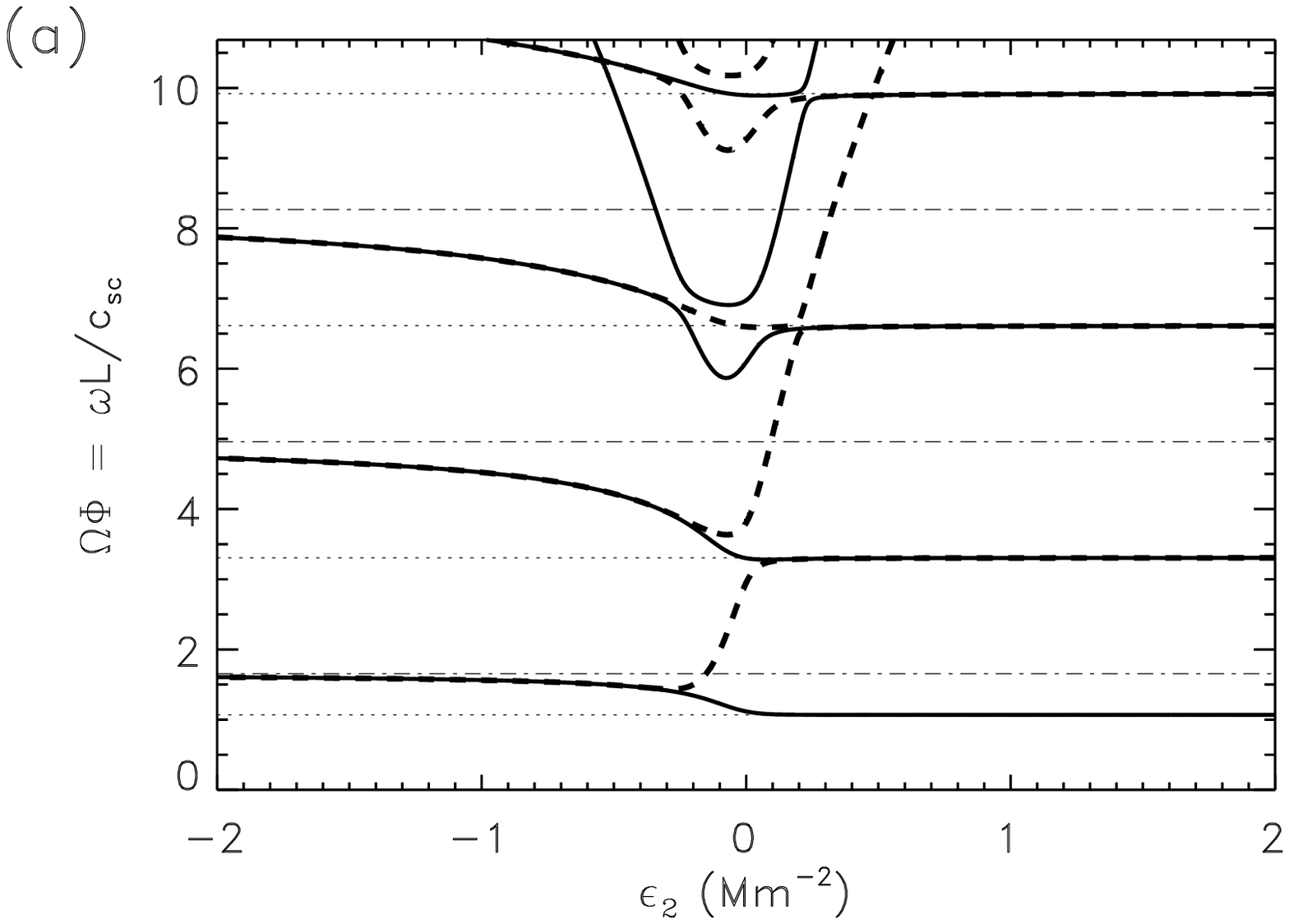}
\includegraphics[width=0.5\textwidth]{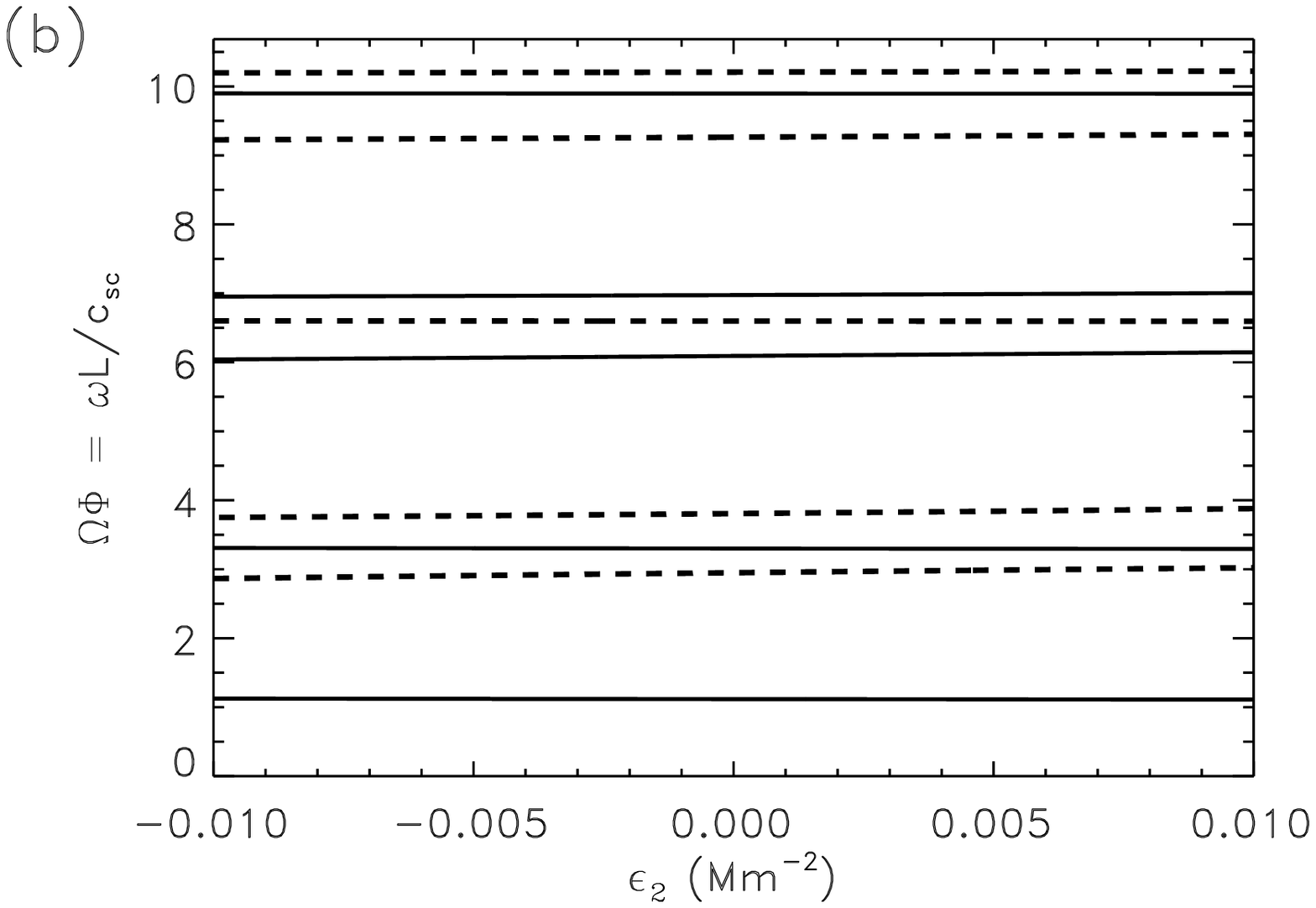}
        \caption{First- and second-kind solutions of the symmetric tube as functions of the expansion parameter $\epsilon_2$. In {\bf (a)} a large range of $\epsilon_2$ values is considered. The dotted lines correspond to the asymptotic values of the frequency for $\epsilon_2 \to \infty$, $\Omega_+$, given by Eq. (\ref{eq:w_asym}). The dot-dashed lines correspond to the asymptotic values of the frequencies for $\epsilon_2 \to -\infty$, $\Omega_-$. In {\bf (b)} only the range of realistic $\epsilon_2$ values is considered. The half-length of the tube is set to $L = 100 \mathrm{~Mm}$, the thread half-length is $l = 5 \mathrm{~Mm}$, the radius of curvature is $R=60 \mathrm{~Mm}$, the coronal sound speed $c_{sc} = 200 \mathrm{~km~s^{-1}}$ , and the temperature contrast $\chi = 100$.}\label{fig:dr_w_modes}
\end{figure}

Figure \ref{fig:dr_w_modes} shows the frequencies of the first-kind (solid lines) and second-kind (dashed lines) modes as a function of the symmetric expansion factor $\epsilon_2$. The fundamental mode is a first-kind solution with a clear dependence on the expansion factor. However, the overtones show a more pronounced dependence on $\epsilon_2$. All modes exhibit interesting behavior for high values of the expansion factor; the modes merge by pairs of different kinds of solutions. For example, the fundamental mode and the first overtone tend to have the same frequency for sufficiently low (negative) values of $\epsilon_2$. Similarly, the branch of the first overtone merges with the branch of the second overtone for sufficiently high values of $\epsilon_2$.

For $\epsilon_2 \gtrsim 0.2 \iiMm$ the mode frequencies approach constant values (see Fig. \ref{fig:dr_w_modes}(a)). For negative values of $\epsilon_2$, the same trend is clear although constant values are not reached in the range of expansion parameter plotted in the figure. The asymptotic behavior of Equations (\ref{eq:dr_sym_tube_1stkind}) and (\ref{eq:dr_sym_tube_2ndkind}) reveals the asymptotic frequencies $\Omega_\pm$ for $\epsilon_2 \to \pm \infty$ as follows:
\begin{equation} \label{eq:w_asym}
\Omega_+= 1, ~ \Omega_\pm=\frac{1}{2}\frac{n_\pm \, \pi }{\Phi (1-W)}~,
\end{equation}
where $\Omega_+= 1$ is the asymptotic value of the fundamental mode for $\epsilon_2 \to + \infty$. In addition, $n_+=2,4,6,\dots$ and $n_-=1,3,5,\dots$ Both $\Omega_+$ and $\Omega_-$ are shown in Figure \ref{fig:dr_w_modes}(a) with horizontal dotted and dot-dashed lines, respectively. From the figure, we see that $\Omega_+= 1$ is exclusive of the fundamental mode. In contrast, the remaining $\Omega_\pm$ values are common for consecutive modes of the first and second kind.

Substituting the dimensional units (Eq. (\ref{dim_units})) in Equation (\ref{eq:w_asym}), we find
\begin{equation} \label{eq:w_asym_dim}
\omega_+= \omega_\mathrm{g}, ~ \omega_\pm=\frac{1}{2}\frac{n_\pm \, \pi c_\mathrm{sc} }{L-l}~.
\end{equation}
These expressions clarify our interpretation of the asymptotic values of the frequencies. For $\epsilon_2 \to \infty$, the fundamental-mode frequency is the pendulum frequency $\omega_\mathrm{g}$. In contrast, the rest of the asymptotic values depend on the sound speed of the hot corona and the lengths of the full tube and the thread. These asymptotic values are the frequencies of a standing slow wave trapped in one of the hot sections of the tube with wavelengths $4(L - l)/n_\pm$. Therefore, for high values of $|\epsilon_2 |$ the entire oscillation in higher order modes is concentrated in the side "coronal" segments of the tube, and the restoring force is the gas pressure gradient. The restoring force of the fundamental mode is gravity for $\epsilon_2 \to \infty$, but for $\epsilon_2 \to -\infty$ the restoring force is the gas pressure gradient.

\subsection{Influence of the expansion parameter $\epsilon_2$ on the oscillation frequency}

\begin{figure}[!h]
\centering\includegraphics[width=9cm]{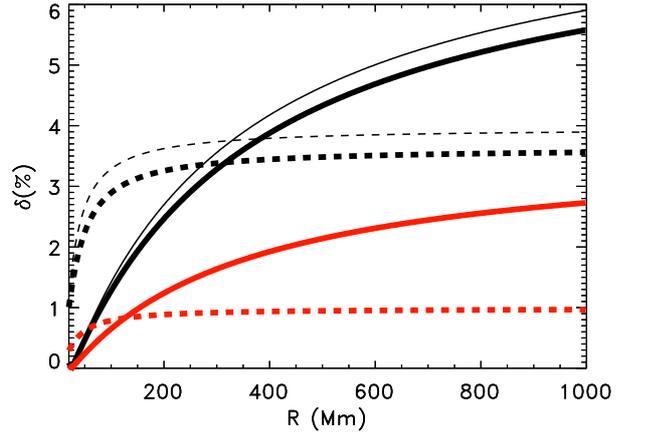}
\caption{Parameter $\delta$ (Eq. \ref{eq:delta}) as a function of radius of curvature R. The black and red lines correspond to $\delta (0, \epsilon_2)$ and $\delta (\epsilon_1, 0)$, respectively. The solid black thick line is $\delta (0, \epsilon_2=0.01 \iiMm)$ whereas the solid black thin line corresponds to $\delta (0, \epsilon_2=-0.01\iiMm)$ for the fundamental mode. The black dashed lines are the same but for the first overtone. The solid red line is $\delta (\epsilon_1=0.1\iMm,0)$ for the fundamental mode and the dashed red line is the same for the first overtone. The remaining parameters are $L = 100 \mathrm{~Mm}$, $l = 5 \mathrm{~Mm}$, $c_{sc} = 200 \mathrm{~km~s^{-1}}$, and $\chi = 100$, as in Fig. \ref{fig:dr_w_modes}.}\label{fig:dependence_R}
\end{figure}

Although the normal modes exhibit interesting behavior for high values of $\epsilon_2$, in prominences the range of values is much smaller: $| \epsilon_2 | \ll 0.01 \iiMm$ as discussed in \S \ref{sec:cross-section-discussion}. In Figure \ref{fig:dr_w_modes}(b) we plotted the frequencies for a realistic range of this expansion parameter. Clearly the mode branches are nearly horizontal lines, indicating that the mode frequencies are nearly constant throughout the realistic range of $\epsilon_2$ values. The reason is that $\gamma_n  \approx \gamma$ in the realistic $\epsilon_2$ range, where the normal modes are similar to those of Paper I. In order to significantly affect the expansion factor it is necessary to have $\gamma_n - \gamma \gg \gamma$, which is equivalent to 
\begin{equation}\label{eq:r-chi-constrain}
\frac{R}{\chi} \gg \frac{\gamma g}{2 c_\mathrm{sc}^2 \epsilon_2}~.
\end{equation}
For a fixed value of $\epsilon_2$ to influence the expansion substantially, a large radius of curvature (very shallow dips) and/or small temperature contrasts (with respect to the adjacent coronal plasma) are required. By defining
\begin{equation}
\epsilon_{2 \mathrm{Lim}}= \frac{\gamma g \chi}{2 c_\mathrm{sc}^2 R}~,
\end{equation}
$\gamma_n$ can be written as
\begin{equation}
\gamma_n=\gamma \left( 1+\frac{\epsilon_2}{\epsilon_{2 \mathrm{Lim}}} \right)~.
\end{equation}
In addition, the effects of tube expansion are greatest for $\epsilon_2 \gg \epsilon_{2 \mathrm{Lim}}$; considering only the range of realistic $\epsilon_2$ values, then $\epsilon_{2 \mathrm{Lim}} \to 0$. This only happens again when $R/\chi \to \infty$, as shown in Equation (\ref{eq:r-chi-constrain}). Using the parameters of Figure \ref{fig:dr_w_modes}, $| \epsilon_{2 \mathrm{Lim}} |= 0.0095 \iiMm$, indicating that $max(|\epsilon_{2}|) \sim |\epsilon_{2 \mathrm{Lim}}|$, where the influence of the expansion on the oscillations is small.  The influence of tube expansion on the mode frequency would be greater if the radius of curvature of the field lines were increased and the temperature contrast, $\chi$, were reduced. The temperature contrast in a prominence is $\sim$100 - 200, so a density contrast of 100 maximizes the influence of the tube expansion. In order to determine the effects of the expansion factors, we compare the mode frequencies with tube expansion, $\omega \, (\epsilon_1, \epsilon_2)$, with the frequencies without expansion, $\omega \, (\epsilon_1=0, \epsilon_2=0)=\omega_0$. We define the relative difference between $\omega_0$ and $\omega$ as 
\begin{equation}\label{eq:delta}
\delta(\epsilon_1, \epsilon_2) =100 \, \left| \frac{\omega-\omega_0}{\omega_0}\right| \% \ .
\end{equation}
In Figure \ref{fig:dependence_R} the black and red lines correspond to $\delta (0, \epsilon_2)$ and $\delta (\epsilon_1, 0)$, respectively; the solid and dashed lines correspond to $\delta$ computed for the fundamental and the first overtone, respectively. In the current section we consider $\delta (0, \epsilon_2)$ (black curves) that for the fundamental modes is below $6\%$ for $ R \le1000 \mathrm{~Mm}$. Similarly, for the two overtones (both black dashed lines) $\delta(0,\epsilon_2)$ is below $\sim 4\%$ for $R \le1000 \mathrm{~Mm}$. Figure \ref{fig:dependence_R} demonstrates that the symmetric expansion factor has negligible influence on thread longitudinal oscillations. For the second and higher overtones, the differences are also constrained to relatively small numbers.

\subsection{Eigenfunctions}\label{sec:eigenfunctions-symtube}

\begin{figure}[!ht]
\begin{center}
\includegraphics[width=0.45\textwidth]{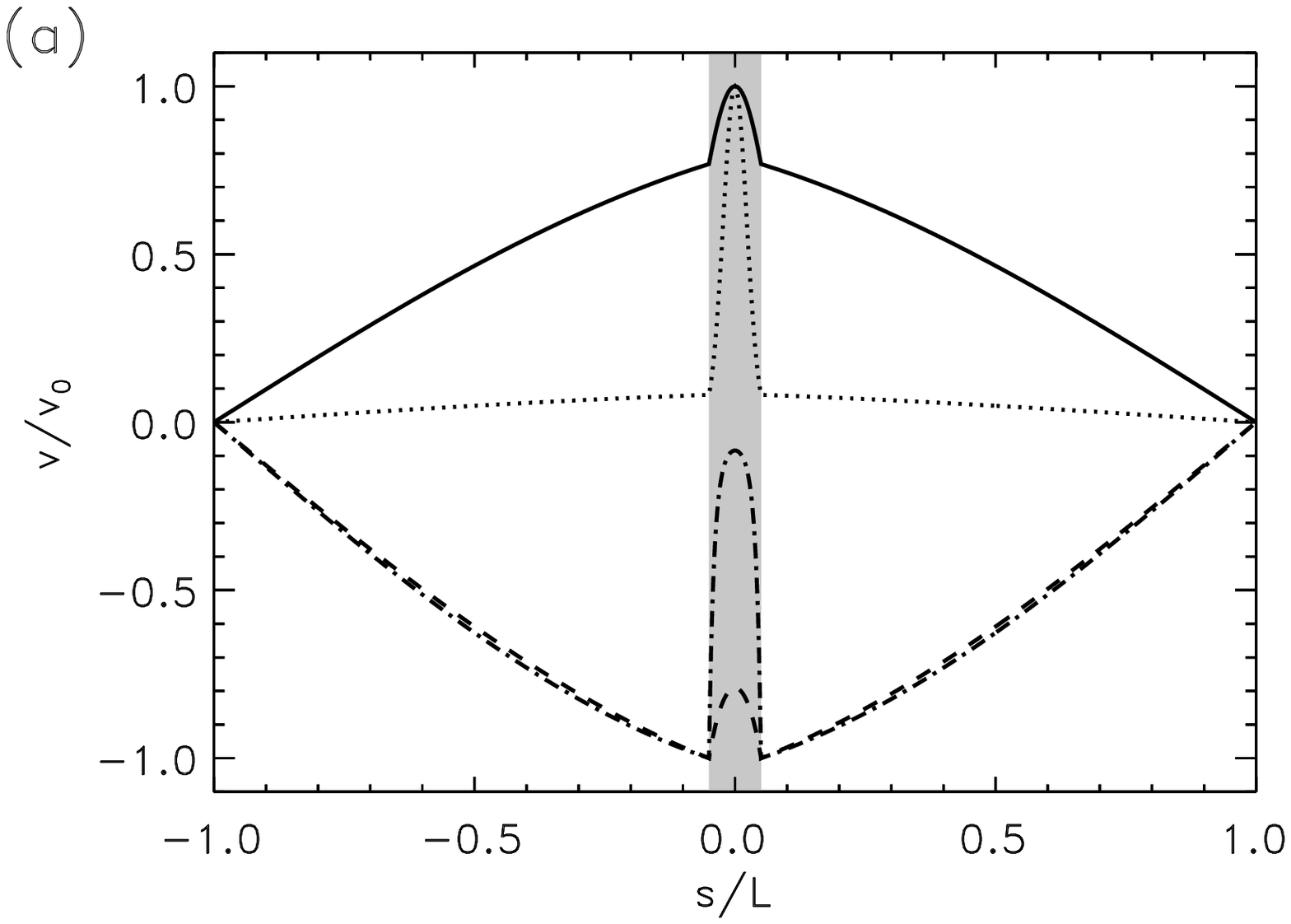}
\includegraphics[width=0.45\textwidth]{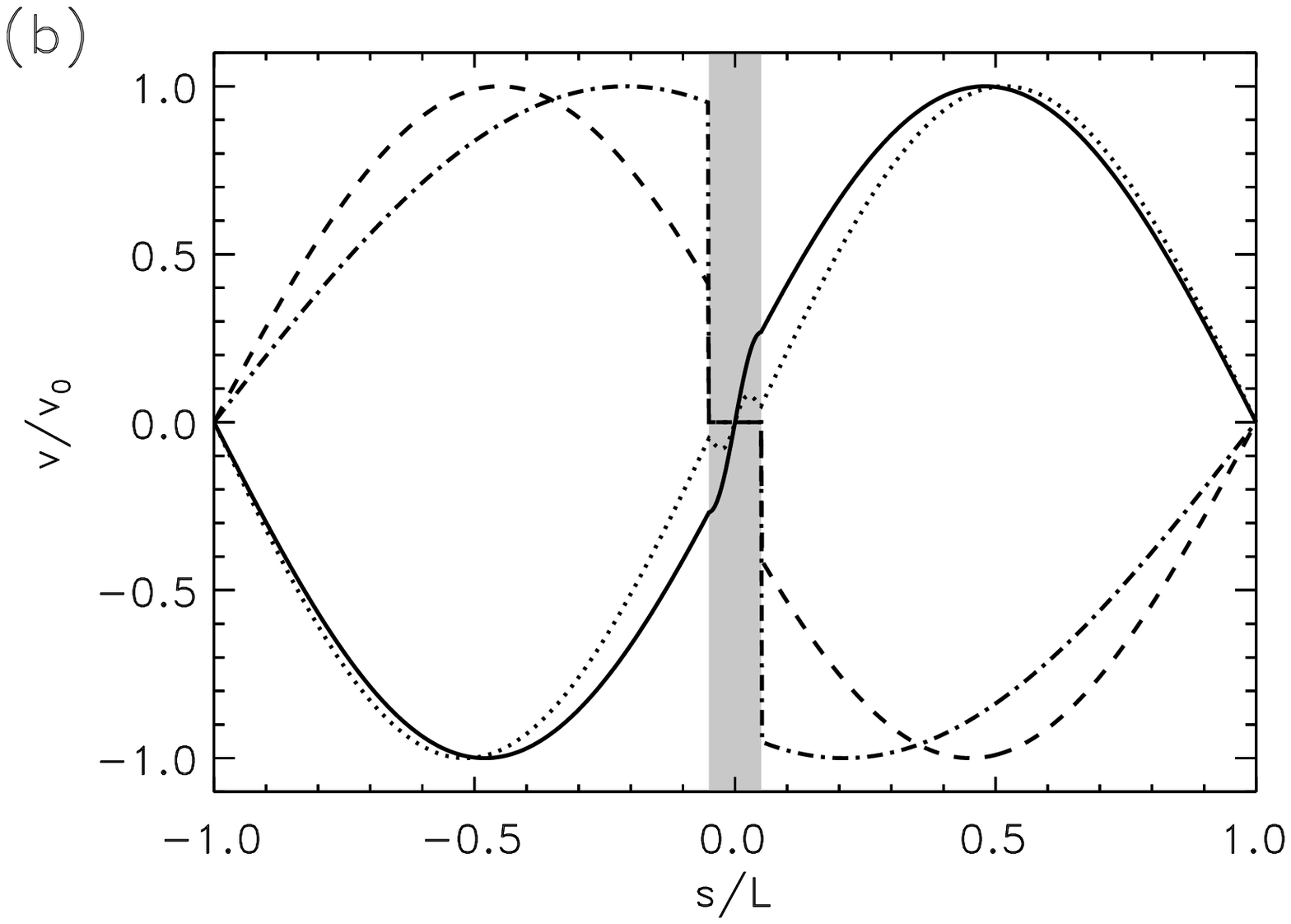}
\end{center}
\caption{Normalized velocities of (a) the fundamental mode (first-kind) for $\epsilon_2=0.01\iiMm$ (solid line), 0.1$\iiMm$ (dotted line), -0.01$\iiMm$ (dashed line), and -0.1$\iiMm$ (dot-dashed line); (b) the first overtone (second-kind) for different expansion parameters. The half-length of the tube is set to $L = 100 \mathrm{~Mm}$, the thread half-length $l = 5 \mathrm{~Mm}$, the radius of curvature $R=60 \mathrm{~Mm}$, the coronal sound speed $c_{sc} = 200 \mathrm{~km~s^{-1}}$ , the contrast $\chi = 100$, and $\epsilon_2=0.005 \iiMm$. The shaded area represents the position of the cool prominence plasma.} \label{fig:v_sym}
\end{figure}

The fundamental first-kind mode has well-defined symmetry with respect to the tube midpoint ($s=0$). For this mode (Fig. \ref{fig:v_sym}(a)) and $\epsilon_2=0.01 \iiMm$ the velocity resembles the results of Paper I: sinusoidal with an increase in the central cool thread. In contrast, for $\epsilon_2=0.1\iiMm$, the velocity is much more concentrated in the thread, falling to less than 5\% of the peak value in the coronal parts of the flux tube.  For higher values of positive $\epsilon_2$ the velocity is entirely concentrated in the thread and negligible in the corona. This is consistent with the asymptotic behavior of the frequencies of Equation \ref{eq:w_asym} where $\Omega_+ \to 1$, indicating that the motion is purely associated with the massive thread. For $\epsilon_2=-0.01\iiMm$ the velocity is also similar to a sinusoidal function but with a decrease of the absolute value at the thread. For more negative values of the parameter, for example, $\epsilon_2=-0.1\iiMm$, this behavior is clearer. The velocity in the cool thread is very low in comparison with the hot parts of the tube. In fact, for even lower values of $\epsilon_2$ the velocity is negligible in the thread. In this case the oscillation velocity is a sinusoidal function in the hot plasma and negligible in the cool thread, for very low values of $\epsilon_2$. This behavior also agrees with the asymptotic values of the frequencies (Eq. \ref{eq:w_asym_dim}).

The first overtone is a second-kind mode with clear antisymmetry with respect to the tube midpoint (Fig. \ref{fig:v_sym}(b)). For $\epsilon_2 = 0.01\iiMm$ the shape of the mode is approximately sinusoidal, except in the thread where the shape is more complex. For high values, for example, $\epsilon_2 = 0.1\iiMm$, the shape in the coronal parts is also similar to a sinusoid but the central segment velocity is low. For increasing values of  $\epsilon_2$ the internal part of the velocity profile decreases; in fact, for sufficiently large  $\epsilon_2$ the velocity in the thread is negligible and the velocity in the corona is a sinusoid. This is consistent with the asymptotic behavior of the frequencies for high values of the expansion parameter. For negative $\epsilon_2$, the central thread velocity also decreases with decreasing expansion parameter and becomes negligible for a sufficiently low value. However, the spatial shape of the velocity in the coronal regions also changes. For $\epsilon_2 = -0.01\iiMm$ the velocity has two lobes with a maximum and a minimum situated at $s/L=\pm0.45,$ respectively. For $\epsilon_2 = -0.1\iiMm$ these positions are $s/L=\pm0.2$. Thus, for decreasing values of $\epsilon_2$,  the positions of the maximum and minimum are closer and the velocity profile resembles the fundamental mode but with opposing velocities on both sides of the thread. For this reason, this mode and the fundamental mode $|v(s)|$ are identical when $\epsilon_2 \to -\infty$. As a result, the fundamental mode and the first overtone have the same frequency for $\epsilon_2 \to -\infty$ (see Fig. \ref{fig:dr_w_modes}(a) and Eq. \ref{eq:w_asym_dim}). Similar behavior can be found for the other overtones, explaining why the overtones have similar frequencies by pairs for high values of $|\epsilon_2|$ (see Fig. \ref{fig:dr_w_modes}(a)).

\section{Asymmetric tubes with $\epsilon_1 \ne 0$ and $\epsilon_2 = 0$}\label{sec:anti_tubes}
\subsection{Eigenvalues}
In this section we explore the normal modes of an asymmetric tube with $\epsilon_1 \neq 0$ and $\epsilon_2 = 0$ (Figs. \ref{fig:cartoon}(c) and \ref{fig:cartoon}(d)). As in \S \ref{sec:sym_tubes}, we apply the boundary conditions (\ref{eq:bc_detailed_interphase}) to find the dispersion relation. In this situation the dispersion relation cannot be factorized in two terms, corresponding to the first- and second-kind solutions. We obtain a very cumbersome dispersion relation, which involves complicated combinations of CHFs. The explicit form of the dispersion relation does not provide relevant information for the purpose of this work and is not shown here.

\begin{figure}[!ht]
\centering\includegraphics[width=0.45\textwidth]{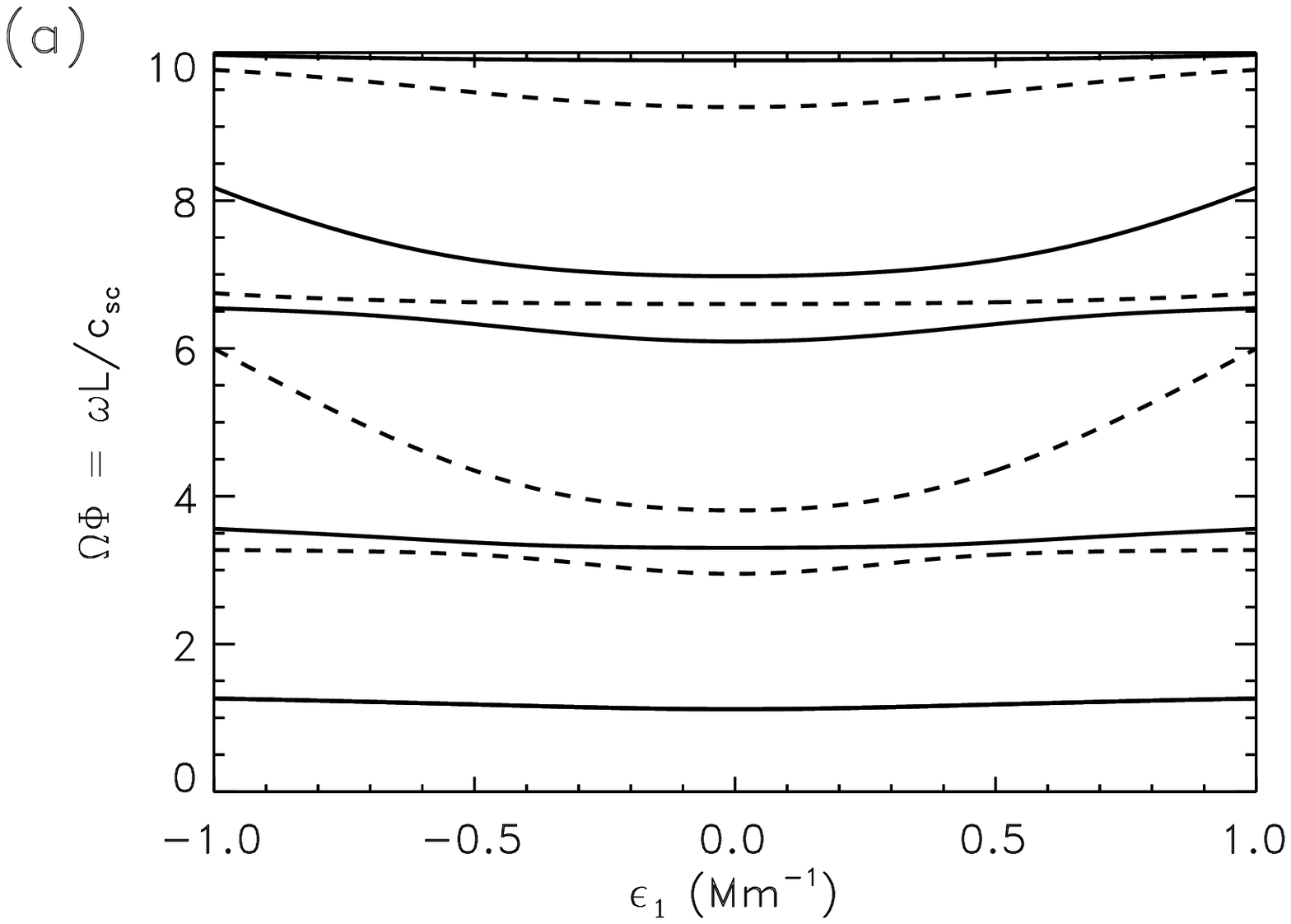}
\centering\includegraphics[width=0.45\textwidth]{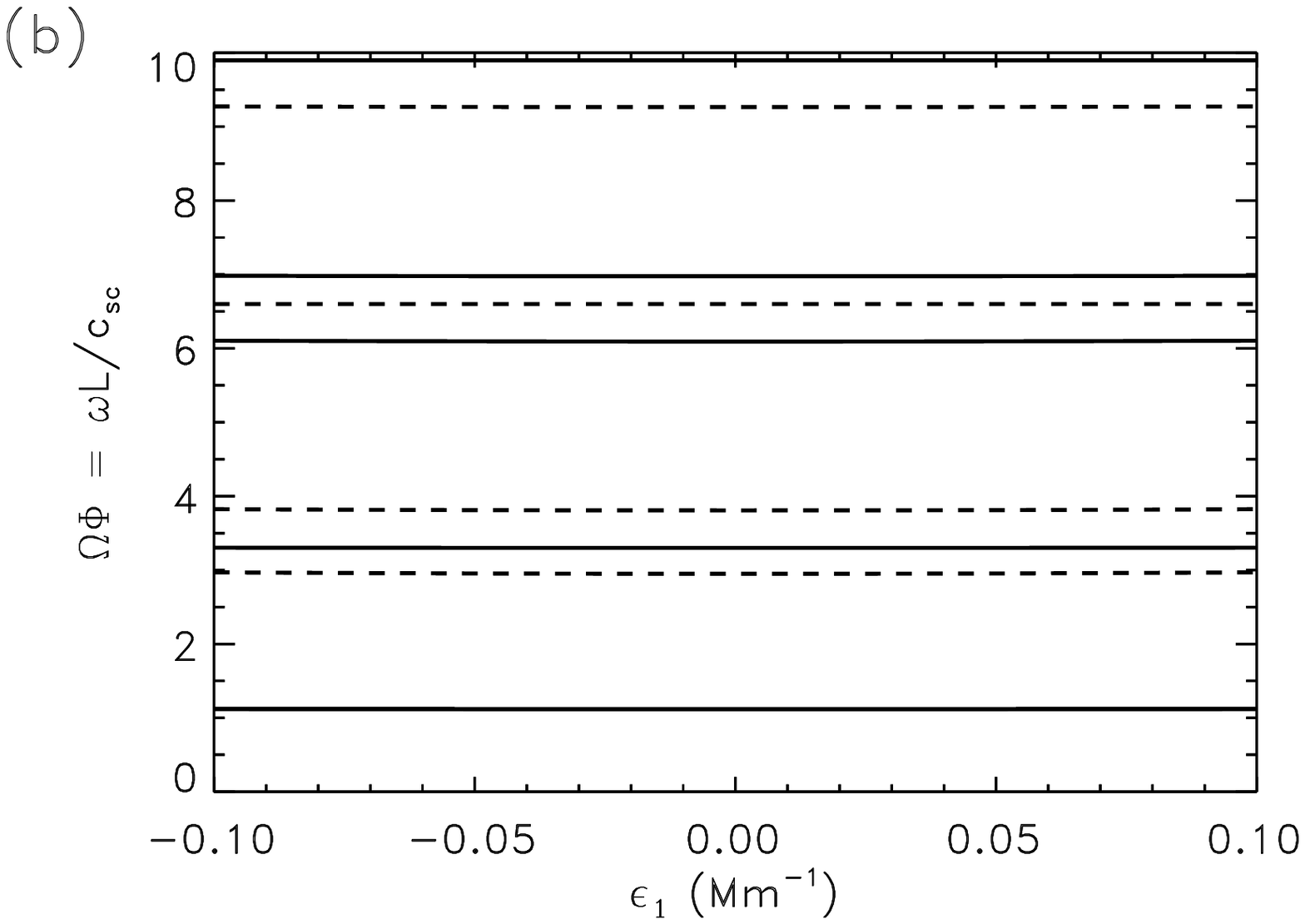}
\caption{First- and second-kind solutions of the asymmetric tube for $\epsilon_1$ and parameters used in Fig. \ref{fig:dr_w_modes}. A large range of values of the expansion parameter $\epsilon_1$ is plotted in {\bf (a)}; only the range of realistic $\epsilon_1$ values is plotted in {\bf (b)}. }
\label{fig:dr_w_modes_atube}
\end{figure}

Figure \ref{fig:dr_w_modes_atube} shows the frequencies from the dispersion relation. Even though the dispersion relation is not factorizable each branch corresponds to first- (solid lines) or second-kind (dashed lines) solutions. The frequencies are symmetric with respect to $\epsilon_1=0$, i.e., $\Omega(\epsilon_1)=\Omega(-\epsilon_1)$, as we expect from arguments given in \S \ref{sec:cross-section-discussion}.

The variation of the frequencies plotted in the figure is small even in the large range of the asymmetric expansion parameter plotted. In Figure \ref{fig:dr_w_modes_atube}(b), it is clear that the variation in the range of realistic values, $|\epsilon_1| \le 0.1 \iMm$, is even smaller. In this panel the different mode branches are almost horizontal lines. For higher values of expansion parameter, $|\epsilon_1| > 1\iMm$, there are avoided crossings between different modes. Similar to the case of Figure \ref{fig:dr_w_modes}, the mode branches tend to finite values asymptotically. However, the values of $\epsilon_1$ where this behavior is clear are very high and clearly out of the range of realistic values of the expansion factor.

\subsection{Influence of the expansion parameter $\epsilon_1$ on the oscillation frequency}

As for the symmetric tube case, the influence of the expansion factor also depends on $R$ and $\chi$. In order to find the normal modes, the boundary conditions (Eq. \ref{eq:bc_detailed_interphase}) are applied to the velocity field (\ref{eq:full_sol_guess}). The spatial dependence of CHF of Equation (\ref{eq:full_sol_guess}) is determined by $\zeta (s)$ (Equation (\ref{eq:zetadef})). When the boundary conditions are applied at $s=\pm l$ the $\zeta$ values are
\begin{equation}
\zeta^2 (s=\pm l) = \frac{1}{2} W^2 \gamma \chi \left( \Phi \pm  \frac{c_\mathrm{s c} \epsilon_1}{W \gamma \chi \omega_g} \right)^2.
\end{equation}
Thus we can define
\begin{equation}
\Phi_n =\Phi \pm  \frac{c_\mathrm{s c} \epsilon_1}{W \gamma \chi \omega_g} = \Phi \left( 1 \pm \frac{\epsilon_1}{\epsilon_{1 \mathrm{Lim}}} \right) ~,
\end{equation}
where
\begin{equation}\label{eq:e1_lim}
\epsilon_{1 \mathrm{Lim}} = \frac{W \gamma \omega_g^2 L \chi}{c_{sc}^2} = \frac{\gamma g}{c_{sc}^2}\frac{l \chi}{R} ~ .
\end{equation}
In order to have an important contribution of the expansion factor, $|\epsilon_1| \gg \epsilon_{1 \mathrm{Lim}}$. In the range of parameters used in Figure \ref{fig:dr_w_modes_atube}, $\epsilon_{1 \mathrm{Lim}} = 0.095\iMm$. Solutions with realistic values of $|\epsilon_1| \le 0.1\iMm$ are not strongly influenced by the expansion. As $\epsilon_1$ is constrained to a finite range then $\epsilon_{1 \mathrm{Lim}} \to 0$, i.e., $l \chi/R \to 0$. This relation indicates that small temperature contrasts, short threads, and relatively flat field lines are necessary for the asymmetric tube expansion to affect the mode frequency significantly. Assuming that the thread half-length is $l = 5 \Mm$ and the minimum temperature contrast is $\chi \sim 100$, we find that $\delta$ is constrained below $3\%$ for $R \le 1000\Mm$ for the fundamental mode (red solid line) and below $1\%$ for the first overtone (red dashed line) (Fig. \ref{fig:dr_w_modes_atube}). We conclude that the influence of the tube expansion is negligible for asymmetric tubes with $\epsilon_2=0$ for both the fundamental mode and higher overtones.

\subsection{Eigenfunctions}
\begin{figure}
\begin{center}
\includegraphics[width=0.45\textwidth]{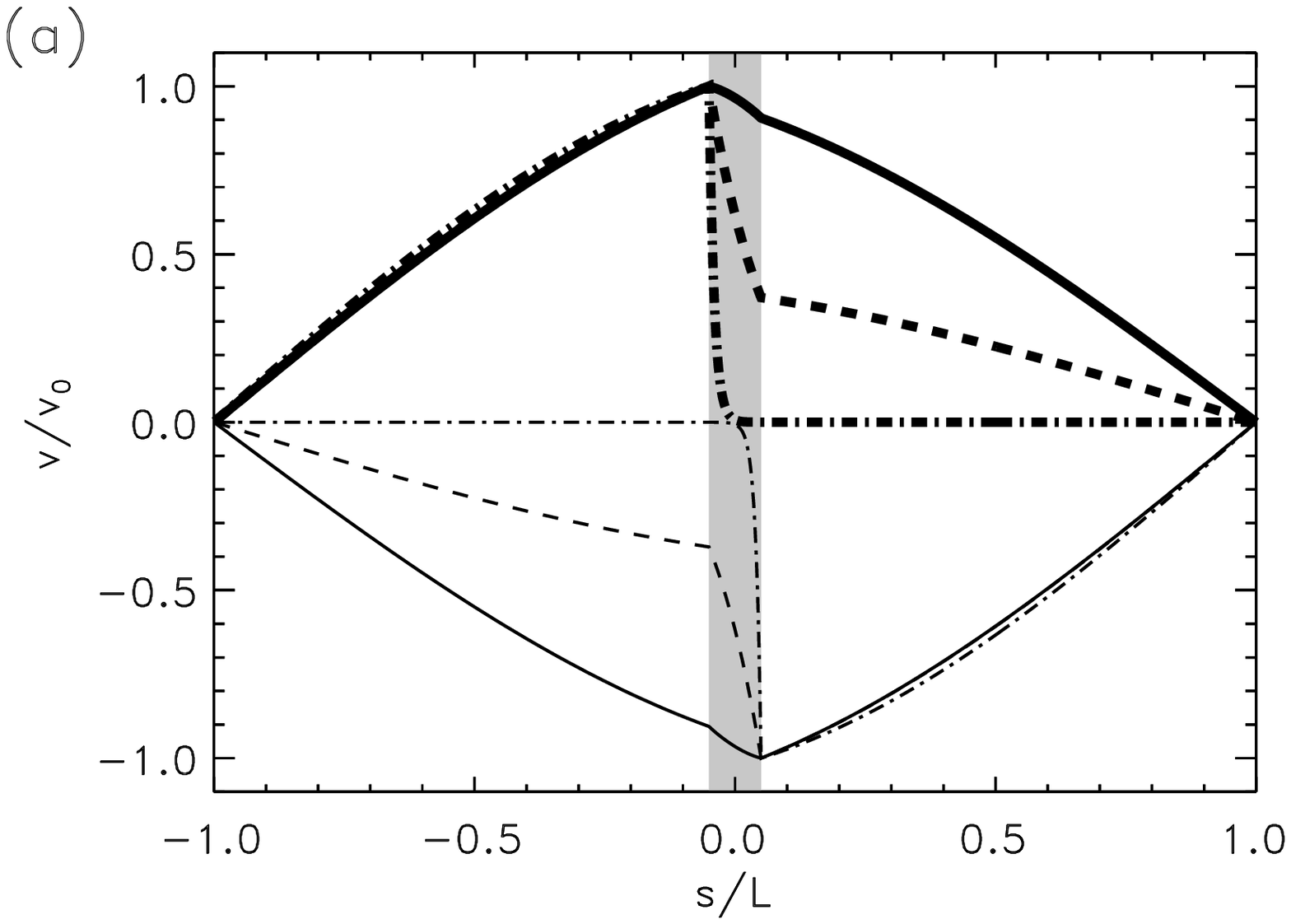}
\includegraphics[width=0.45\textwidth]{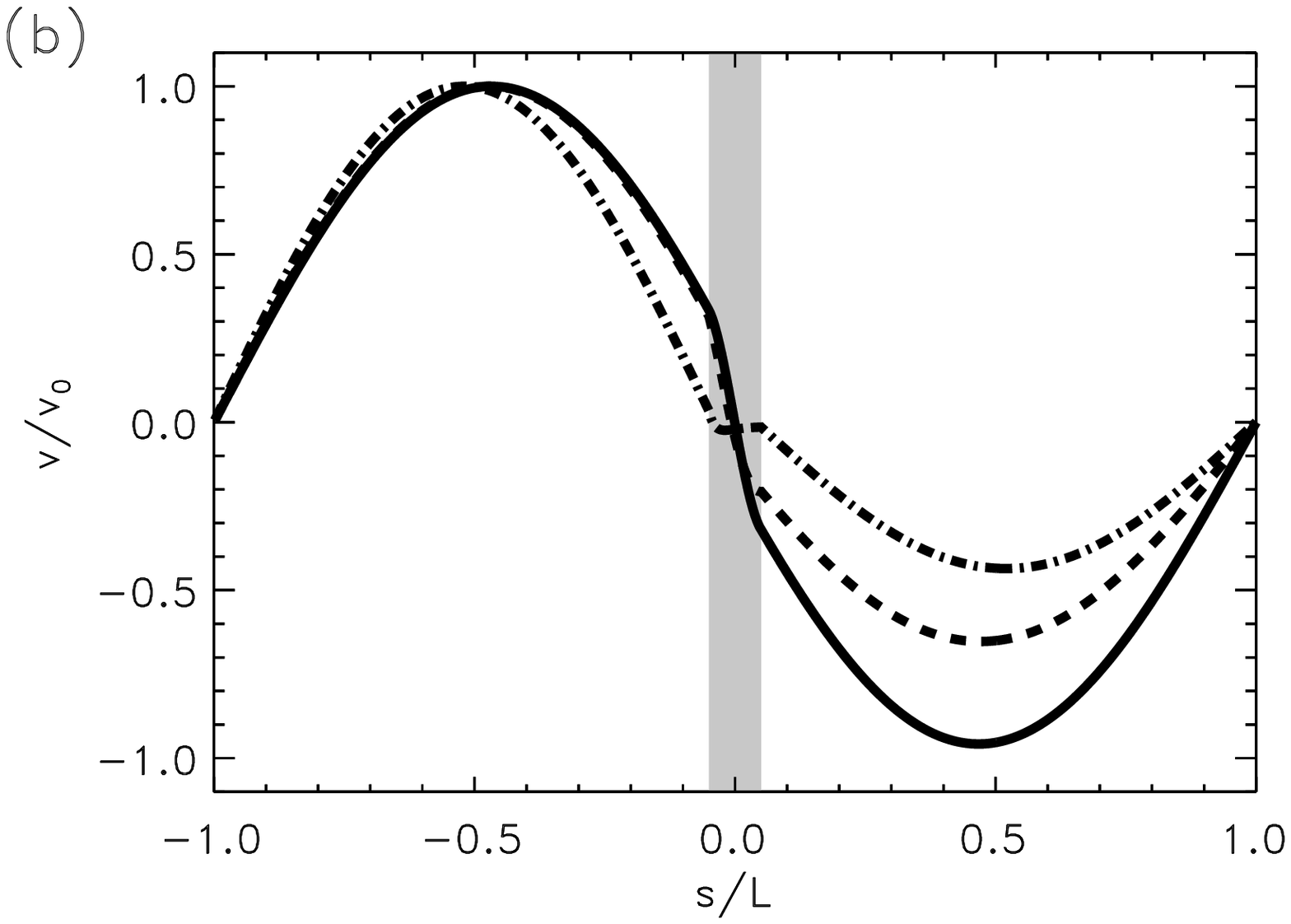}
\caption{Plot of the first- and second-kind solutions for the asymmetric tube. In (a) the thick lines correspond to first-kind solutions with positive expansion parameters $\epsilon_1=0.01\iMm$ (solid line), $\epsilon_1=0.1\iMm$ (dashed line), and $\epsilon_1=1\iMm$ (dot-dashed line). The thin lines correspond to negative expansion parameters $\epsilon_1=-0.01\iMm$ (solid line), $\epsilon_1=-0.1\iMm$ (dashed line), and $\epsilon_1=-1\iMm$ (dot-dashed line). In this plot it is clear that the solutions with $\epsilon_1$ positive or negative are identical under the inversion of the coordinate, $s$ to $-s$. Similarly, in (b) the second-kind solutions are plotted. For simplicity, in (b) only positive expansion parameters solutions are plotted.}\label{fig:v_antisym}
\end{center}
\end{figure}

Figure \ref{fig:v_antisym} clearly shows that the first- and second-kind normal-mode solutions are asymmetric with respect to $s=0$, and this figure reveals the character of these modes in the limit $\epsilon_1 \to 0$. First-kind solutions have the same sign along $s,$ whereas the second-kind have a node in the cool thread (not at $s=0$) where the solution changes sign. In Figure \ref{fig:v_antisym}(a) we plotted solutions for positive and negative $\epsilon_1$. Both solutions are equivalent to flipping the tube by exchanging $s$ with $-s$. In both kinds, the solutions are more concentrated on one side of the tube for larger expansion parameters. For the first kind and $\epsilon_1 = 1\iMm$, there is no motion on the right side ($s\ge+l$) of the flux tube and all motion is concentrated on the left side. For $\epsilon_1 = -1\iMm$, the opposite happens and all motion is concentrated on the right side of the tube. For the second-kind solutions, the motion persists in both sides of the tube; however, for $\epsilon_1 = 1\iMm$, the motion is concentrated on the left side and the velocity in the thread (shaded area) is almost zero.

Although the mode frequencies are not significantly affected by realistic values of the tube expansion, the normal mode shapes are clearly influenced by the asymmetric expansion factor (Figure \ref{fig:v_antisym}).

\section{The general case with $\epsilon_1 \ne 0$ and $\epsilon_2\ne0$}\label{sec:general_case}
In a realistic situation, the shape of a flux tube is given by both nonzero symmetric and asymmetric expansion factors (see Fig. \ref{fig:epsilons}). In this Section, therefore, we consider this general case with $\epsilon_1 \ne 0$ and $\epsilon_2 \ne 0$. As in the previous Section, the resulting dispersion relation is cumbersome and is not shown here.

\begin{figure}
\includegraphics[width=0.5\textwidth]{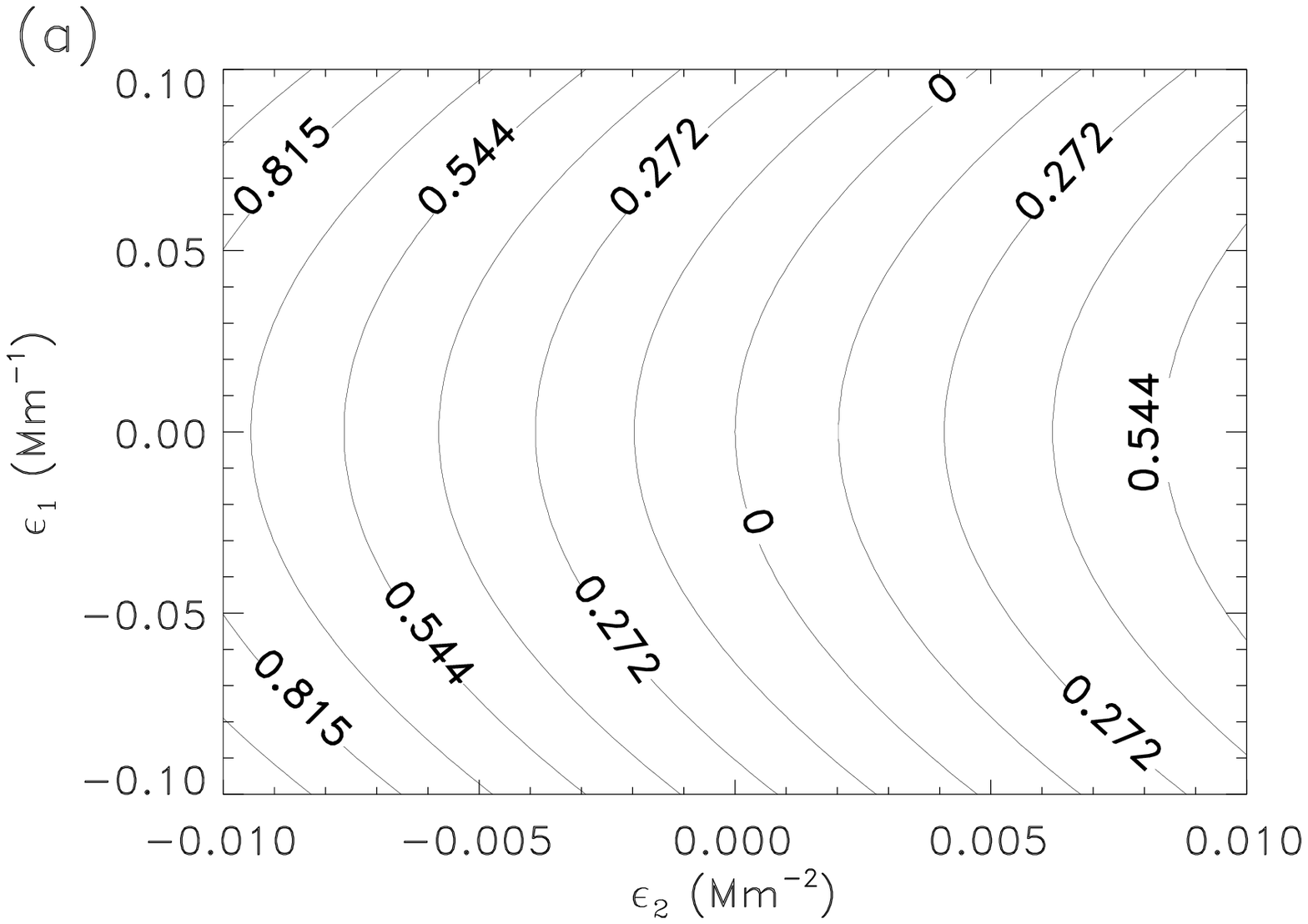}\\
\includegraphics[width=0.5\textwidth]{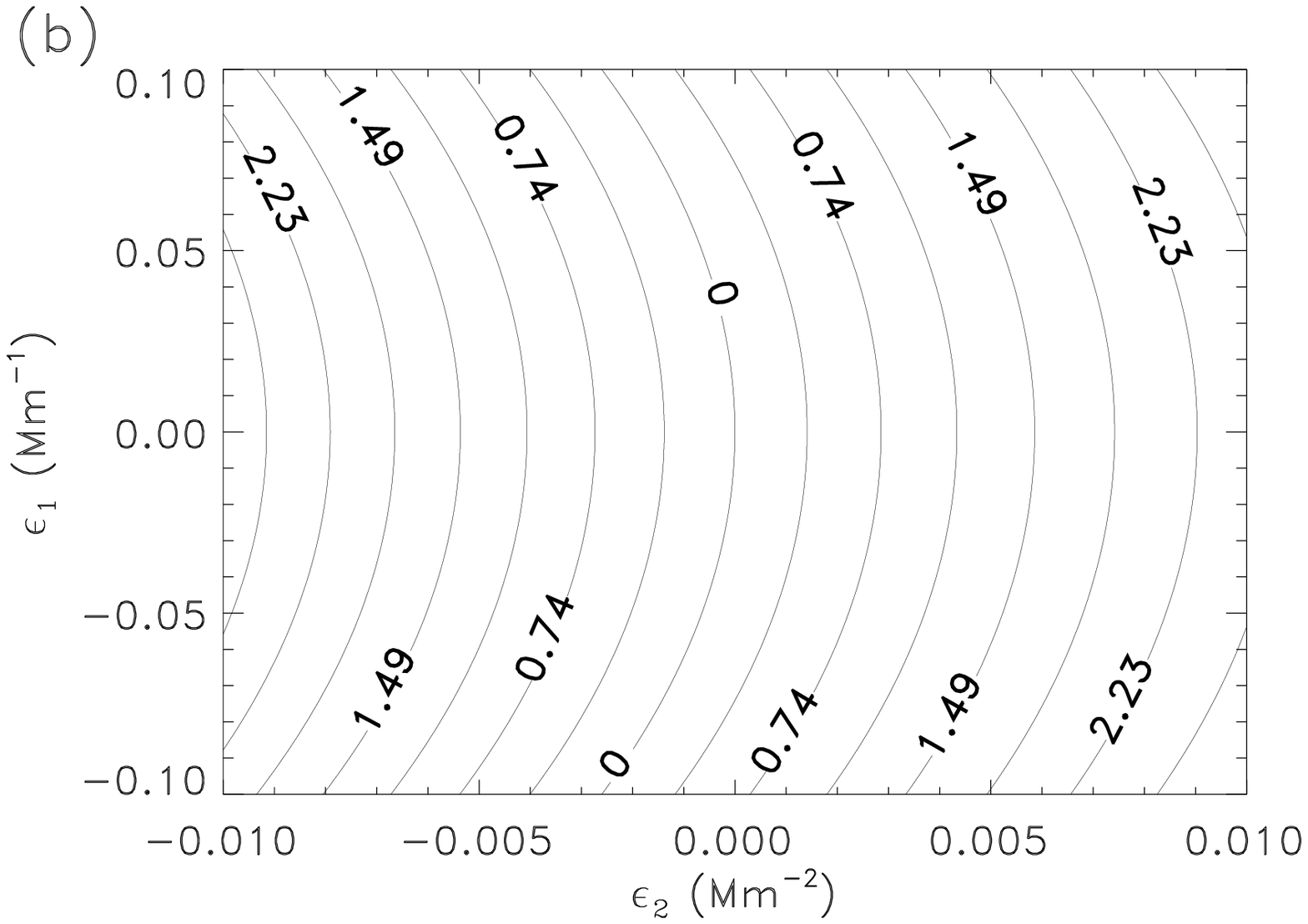}
\caption{Plot of the contour lines of the $\delta$ function of (a) the fundamental mode and (b) the first overtone. The  $\delta$ function is computed with Equation (\ref{eq:delta}) using the mode frequencies for realistic $\epsilon_1$ and $\epsilon_2$ values. The labels of the contour lines are a percentage of the reference $\omega_0$ values. }
\label{fig:deltabarrido}
\end{figure}

\begin{figure}
\includegraphics[width=0.5\textwidth]{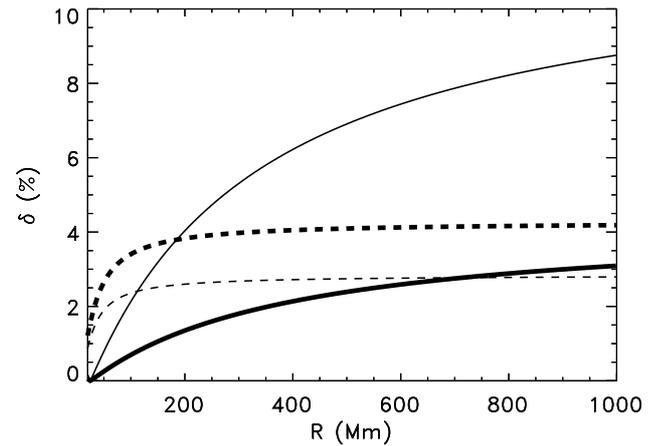}
\caption{As in Fig. \ref{fig:dependence_R}. The solid black thick line is $\delta(\epsilon_1=0.1\iMm$, $\epsilon_2=0.01\iiMm),$ whereas the solid black thin line is $\delta(\epsilon_1=0. 1\iMm$, $\epsilon_2=-0.01\iiMm)$, both for the fundamental mode. The dashed lines are similar but for the first overtone. The remaining parameters are $L = 100 \mathrm{~Mm}$, $l = 5 \mathrm{~Mm}$, $c_{sc} = 200 \mathrm{~km~s^{-1}}$, and $\chi = 100$, as in Fig. \ref{fig:dr_w_modes}.}\label{fig:differences_combined}
\end{figure}

Figure \ref{fig:deltabarrido} shows the $\delta$ function of Equation (\ref{eq:delta}) for the fundamental mode (Fig. \ref{fig:deltabarrido}(a)) and for the first overtone (Figure \ref{fig:deltabarrido}(b)) for realistic values of $\epsilon_1$ and $\epsilon_2$. For the fundamental (first overtone) mode, the differences in $\delta$ between this case and the tube without expansion are less than $1\%$ ($3\%$). Therefore, the influence of the combined expansion terms on the fundamental and higher order modes is negligible for the general case. The combined influence of both expansion factors also depends on the parameter $R/\chi$, as in the symmetric or asymmetric tube cases. Figure \ref{fig:differences_combined} shows $\delta(\epsilon_1=0.1\iMm, \epsilon_2=\pm 0.01\iiMm)$ for the frequencies of the fundamental and the first overtone modes as functions of $R$, assuming the minimum value of $\chi=100$.  Here the difference of both $\delta(\epsilon_1,\epsilon_2=\pm 0.01\iiMm)$ is now larger than that shown in Figure \ref{fig:dependence_R}. The case of $\delta(\epsilon_1=0.1\iMm, \epsilon_2=0.01\iMm) < 3\%$ for $R\le 1000 ~\Mm$. In contrast, $\delta(\epsilon_1=0.1\iMm, \epsilon_2=-0.01\iiMm)$ reaches almost $9\%$ at $R= 1000~ \Mm$. Thus, the combined action of the symmetric and asymmetric expansion factors increases the difference with respect to the uniform tube for $\epsilon_1=0.1\iMm$ and $\epsilon_2=-0.01\iiMm$, and decreases it for $\epsilon_1=0.1\iMm$ and $\epsilon_2=0.01\iiMm$. The full realistic range of the expansion parameters is shown in Figure \ref{fig:deltabarrido}; lower values produce smaller differences from the uniform-tube case.

\begin{figure}
\includegraphics[width=0.5\textwidth]{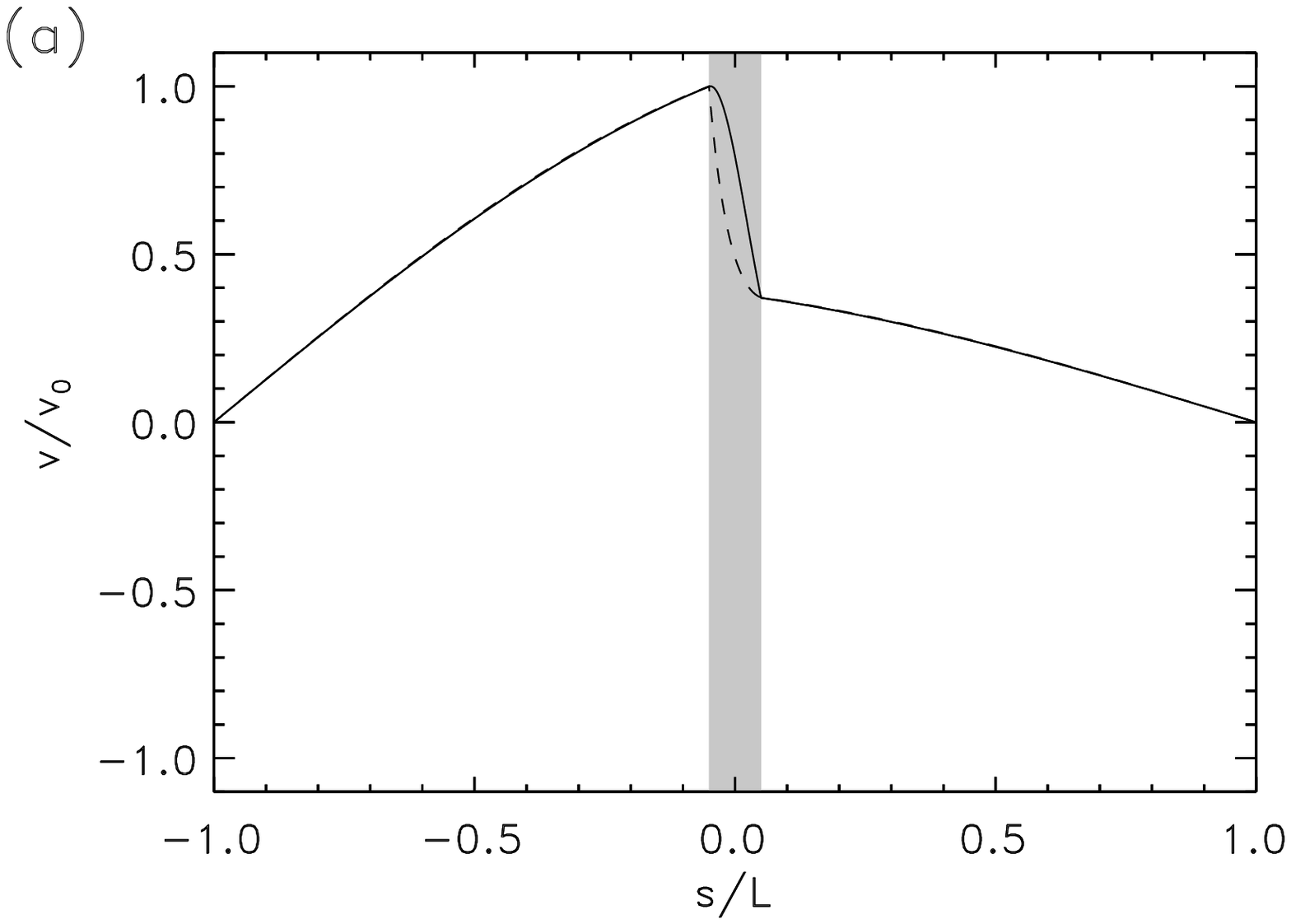}\\
\includegraphics[width=0.5\textwidth]{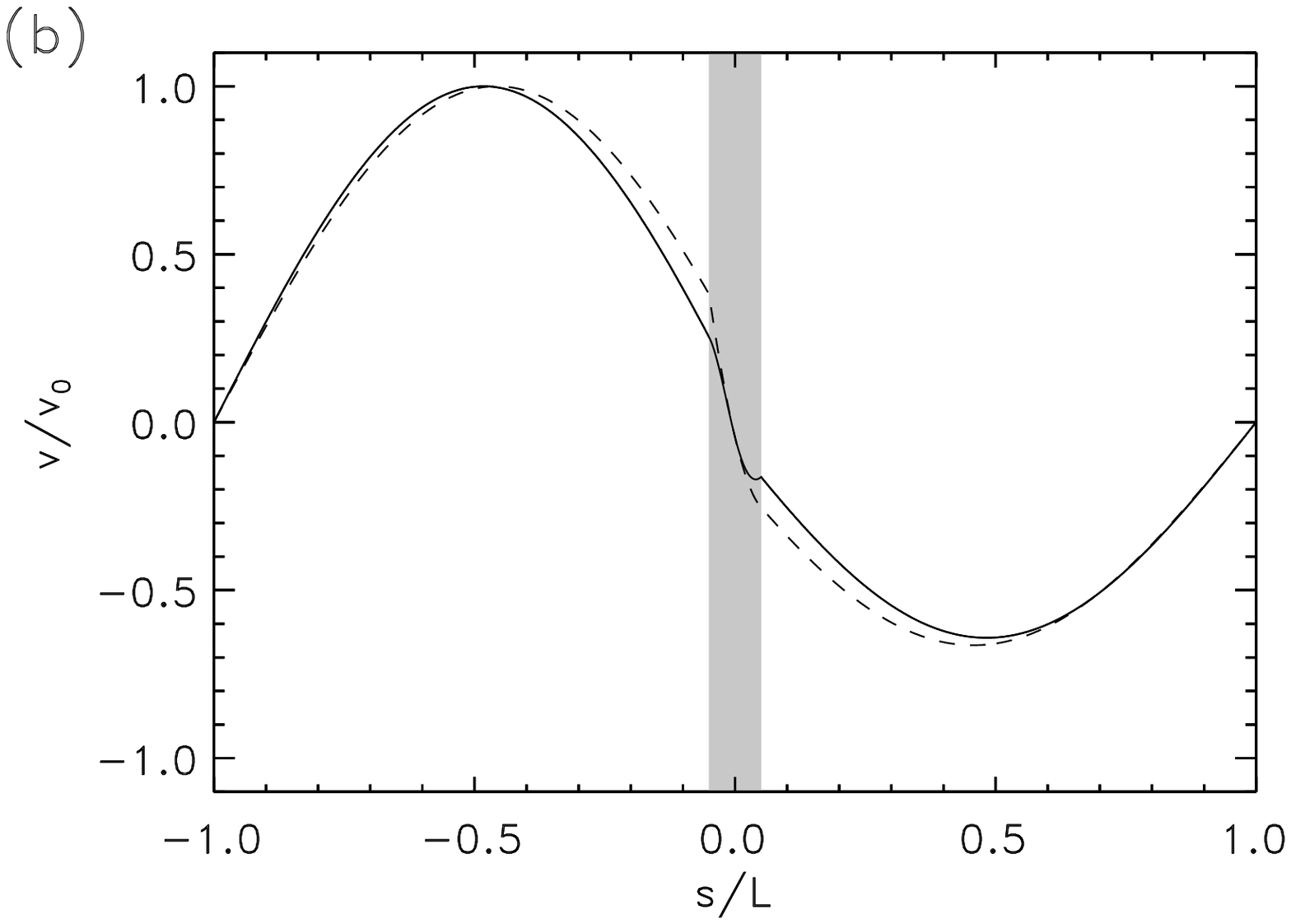}
\caption{Plot of the fundamental mode (a) and the first overtone (b) solutions for $\epsilon_1=0.1 \iMm$ and $\epsilon_2=0.01\iiMm$ (solid line) and $\epsilon_1=0.1 \iMm$ and $\epsilon_2=-0.01\iiMm$ (dashed lines) for $L = 100 \mathrm{~Mm}$, $l = 5 \mathrm{~Mm}$, $R=60 \mathrm{~Mm}$, $c_{sc} = 200 \mathrm{~km s^{-1}}$, and $\chi=100$.}\label{fig:v_both}
\end{figure}

Figure \ref{fig:v_both} shows the fundamental and first overtone solutions corresponding to the extreme cases: $\epsilon_1=0.1\iMm$ with $\epsilon_2=\pm 0.01\iiMm$.

For the fundamental mode (Fig. \ref{fig:v_both}(a)) the solutions coincide in the coronal parts of the tube. In the cool parts (shaded area) the differences are clear: higher velocities for $\epsilon_2=0.01 \iiMm$ and slightly lower velocities for $\epsilon_2=-0.01\iiMm$. The asymmetric term influences the shape of the fundamental mode, as in the case shown in Figure \ref{fig:v_antisym}(a). The results of Section \ref{sec:anti_tubes} remain valid here. Even though the influence of the expansion terms on the frequencies is small, the asymmetric term introduces a difference in the shape of the normal mode, such that both ends of the thread oscillate with significantly different velocities (by a factor of $\sim$3). Similarly, for the first overtone (Fig. \ref{fig:v_both}(b)) the two extremes of $\epsilon_2$ produce only minor differences in the velocity profile, while the $\epsilon_1$ term introduces a clear asymmetry in the velocity.

\section{Discussion and conclusions}\label{sec:conclusions}

In this work we investigated the influence of the cross-sectional area, i.e., magnetic field, variations on the frequencies and velocity profiles of the normal modes in a flux tube containing a prominence thread surrounded by hot corona. We characterized the area variations with two parameters: the symmetric and asymmetric terms that we called expansion parameters. The asymmetric term produces flux tubes with cross-sections that are not symmetric about the tube midpoint, i.e., with one end of the tube that is thicker than the other. The symmetric term produces tubes with symmetric cross-sections, where both ends have the same area; in this case the midpoint of the tube is thinner than the ends for positive values and thicker for negative values of the symmetric expansion factor. Our model prominence flux tube is segmented into three isothermal regions: a cool thread with a symmetric or asymmetric area expansion in the center, bounded by two horizontal side segments of hot coronal plasma with no area expansion. 

We derived the equation that describes the linear perturbations of the velocity along the model flux tube, and solved for the normal modes for symmetric, asymmetric, and mixed flux-tube geometries. In general, the normal modes lack symmetry with respect to the tube midpoint and can be classified into two categories. The first-kind modes have an even number of nodes (zeroes) of the velocity along the entire tube, whereas the second-kind have an odd number of nodes. The fundamental mode, which corresponds to the pendulum mode of Paper I, is a first-kind mode in all cases considered. 

We found that the expansion factors generally exert little influence on the longitudinal oscillation frequencies.  For small radius of curvature and large temperature contrast, the influence of the cross-sectional variation increases although not significantly. In a tube described by a combination of the symmetric and asymmetric expansion factors the influence is in some cases slightly larger. For example, in the limit of almost straight field lines (large $R$), the frequencies diverge by $<9\%$ from the uniform-tube frequencies. The shape of the flux tube also affects the spatial profile of the normal-mode velocity. For the fundamental mode, the symmetric expansion factor produces a relatively small enhancement of the velocity in the center of the tube. In contrast, the asymmetric expansion term clearly produces different velocities at the thread ends by as much as a factor of 3 for realistic expansion factors. This is the most significant result in terms of observable signatures that could potentially provide insight into the geometry of prominence-bearing flux tubes. 

We conclude that, in general, the longitudinal oscillations are not strongly influenced by nonuniform magnetic field strength, i.e., cross-sectional area, along the flux tube. Relatively high values of the asymmetric expansion term, however, can produce significant velocity differences at both ends of the prominence threads. These results confirm that the pendulum model of Paper I remains valid for flux tubes with a range of cross-sectional area variations that are consistent with observations, but the model considered here does not allow interactions between the plasma and magnetic field.  Recently we simulated the effects of perturbing localized cool plasma supported by a 2D dipped magnetic field \citep{luna2016}. We found that the back-reaction of the field to the plasma oscillation is very small, validating the simpler assumption of rigid flux tubes in Paper I and the present study. Nonlinear 3D simulations \citep[e.g.,][]{xia2016} are needed to fully understand the oscillatory modes of prominence plasma embedded in the hot, magnetized corona, particularly for the large-amplitude longitudinal oscillations commonly observed near energetic events \citep[e.g.,][]{luna2014}.

\begin{acknowledgements} M. Luna acknowledges the support by the Spanish Ministry of Economy and Competitiveness through projects AYA2011-24808, AYA2010-18029, and AYA2014-55078-P. This work contributes to the deliverables identified in FP7 European Research Council grant agreement 277829, ``Magnetic Connectivity through the Solar Partially Ionized Atmosphere'' (PI: E. Khomenko). J. T. acknowledges support from the Spanish ``Ministerio de Educaci{\'o}n y Ciencia'' through a Ram{\'o}n y Cajal grant and support from MINECO and FEDER funds through project AYA2014-54485-P. M. L., J. T., and J. K. acknowledge support from the International Space Science Institute (ISSI) to the Team 314 on ``Large-Amplitude Oscillation in prominences'' led by M. Luna.
\end{acknowledgements}

\bibliographystyle{aa} 
\bibliography{wavesgravity} 
\end{document}